\documentclass[pra,twocolumn,superscriptaddress,notitlepage]{revtex4-1}
\usepackage{graphicx,amsmath,amsfonts,amssymb,upgreek,txfonts,color}
\usepackage[colorlinks,linkcolor=blue,citecolor=red,urlcolor=blue,breaklinks=true]{hyperref}
\begin{document}

\title{Controllable generation of photons and phonons in a coupled BEC-optomechanical-cavity via the parametric dynamical Casimir effect}
\author{Ali Motazedifard} 
\email{motazedifard.ali@gmail.com}
\address{ Department of Physics, Faculty of Science, University of Isfahan, Hezar Jerib, 81746-73441, Isfahan, Iran}
\address{ School of Physics, Institute for Research in Fundamental Sciences (IPM), 19395-5531, Tehran, Iran}

\author{A. Dalafi} 
\email{a\_dalafi@sbu.ac.ir}
\address{Laser and Plasma Research Institute, Shahid Beheshti University, Tehran 19839-69411, Iran}

\author{M. H. Naderi} 
\email{mhnaderi@sci.ui.ac.ir}
\address{ Department of Physics, Faculty of Science, University of Isfahan, Hezar Jerib, 81746-73441, Isfahan, Iran}
\address{Quantum Optics Group, Department of Physics, Faculty of Science, University of Isfahan, Hezar Jerib, 81746-73441, Isfahan, Iran}

\author{R. Roknizadeh}
\email{rokni@sci.ui.ac.ir}
\address{ Department of Physics, Faculty of Science, University of Isfahan, Hezar Jerib, 81746-73441, Isfahan, Iran}
\address{Quantum Optics Group, Department of Physics, Faculty of Science, University of Isfahan, Hezar Jerib, 81746-73441, Isfahan, Iran}


\date{\today}
\begin{abstract}
We theoretically propose and investigate a feasible experimental scheme for the realization of the dynamical Casimir effect (DCE) in a hybrid optomechanical cavity with a moving end mirror containing an interacting cigar-shaped Bose-Einstein condensate (BEC). 
We show that in the red-detuned regime of cavity optomechanics together with the weak optomechanical coupling limit by \textit{coherent} modulation of the \textit{s}-wave scattering frequency of the BEC and the mechanical spring coefficient of the mechanical oscillator (MO), the mechanical and atomic quantum vacuum fluctuations are parametrically amplified, which consequently lead to the generation of the mechanical/Bogoliubov-type Casimir phonons. 
Interestingly, in the coherent regime corresponding to the case of largely different optomechanical coupling strengths of the cavity field to the BEC and the MO, or equivalently largely different cooperativities, one can generate a large number of Casimir photons due to the amplification of the intracavity vacuum fluctuations \textit{induced} by the time modulations of the BEC and the MO. 
The number of generated Casimir particles are externally controllable by the cooperativities, and the modulation amplitudes of the atomic collisions rate and the mechanical spring coefficient.

\end{abstract}

\pacs{, , , }

\maketitle

\section{Introduction}
One of the most distinctive features of the quantum field theory is the existence of quantum vacuum fluctuations that have no counterpart in the classical physics. Generation of real particles out of the quantum vacuum fluctuations, or dynamical vacuum amplification, is one of the manifestations of such fluctuations on the macroscopic level. 
Examples of the quantum vacuum amplification phenomenon include the \textit{Schwinger} effect \cite{Schwinger}, which is the electron-positron pair production from the vacuum under the action of strong electric fields; 
\textit{Hawking} radiation \cite{Hawking}, which results from a vacuum instability of quantum fields at a black-hole horizon; 
the \textit{Unruh} effect \cite{Unruh}, which predicts that an accelerating observer traveling through the Minkowski vacuum will observe a thermal spectrum of particle excitations; 
and the \textit{dynamical} Casimir effect (DCE) which concerns the generation of real particles (conventionally creation of photons) out of the quantum vacuum \cite{Yablonovitch,Schwinger2,Wilson} when the boundary conditions of the field are varied at a fast-enough rate\cite{Moore,Davies,Dodonov1,Nation}.

In spite of the \textit{static} Casimir effect which results from a mismatch of vacuum modes in the space domain, the DCE arises from a mismatch of vacuum modes in the time domain.
The DCE can be explained qualitatively as a particular kind of the parametric amplification of quantum vacuum fluctuations in systems with time-dependent parameters leading to photon generation. 
In addition to the theoretical investigations on the issue of particle generation via the DCE in a large variety of systems, ranging from cosmology to non-stationary cavity QED \cite{Daviesbook.Dodonov2.Dalvit.Dodonov3.Crocce.Dodonov4.Dodonov5}, various theoretical schemes for practical applications of the DCE have been suggested, including generation of photons with nonclassical properties \cite{Dodonov6,Dodonov7,Johansson}, generation of atomic squeezed sates \cite{Bhattacherjee}, generation of multipartite entanglement in cavity networks \cite{Felicetti}, and generation of EPR quantum steering and Gaussian interferometric power \cite{Sabin}. 

Most of the schemes proposed until now to realize the DCE can be roughly separated into two categories: (a) the schemes based on the real mechanical motion of boundaries, e.g., mirrors of a cavity, a mechanism referred to as  motion-induced DCE (MIDCE) in the literature \cite{Barton}; and (b) the schemes based on the parametric amplification of vacuum fluctuations in media without moving boundaries, a process which is a kind of imitation of boundary motion and  known as parametric DCE (PDCE) \cite{DodonovPDCE}.
 
From the experimental point of view, in order to generate a measurable flux of real photons from vacuum, the real moving boundaries should oscillate at very high frequencies. Indeed, the resonance condition for realization of the MIDCE requires the mechanical frequency to be at least twice that of the cavity which still remains a serious problem. 
Recently a high mechanical frequency as large as $ \omega_m/2\pi \sim 6 $GHz has been reported \cite{high mechanical frequency}. However, for the generation of Casimir radiation at the frequency of about $ \omega_c/2\pi = 5 $GHz a still higher mechanical frequency is needed. Consequently, alternative schemes based on imitation of boundary motions (PDCE) have been proposed. Some examples include periodic modulation of the optical properties of the boundary \cite{Lombardi,Dodonov8,Dodonov9} or of the optical path length of a cavity \cite{Dezael,Faccio,Motazedifard DCE}, and enhancing Casimir photon generation in a cavity within the driven Rabi model of the qubit-field interaction with a time-dependent modulation \cite{Hoeb}. 
Some other experimental schemes aiming to observe the DCE can be found in \cite{Lombardi,Agnesi,Kawakubo}. Recently, the DCE has been realized experimentally in superconducting circuit QED through fast-modulating either the electrical boundary condition of a transmission line \cite{Pourkabirian} or the effective speed of light in a Josephson metamaterial \cite{Lahteenmaki}. 

Besides, the MIDCE or PDCE proposals include analog models for generation of other particles than photons mostly in the physical systems including Bose-Einstein condensates (BECs). For example, the DCE of phonons in a time modulated atomic BEC \cite{Recati,Jaskula}, dynamical Casimir emission of Bogoliubov excitations in an exciton-polariton condensate suddenly generated by an ultrashort laser pulse \cite{Koghee}, phononic DCE in a time-modulated quantum fluid of light \cite{Busch}, DCE of magnon excitation in a spinor BEC driven by a time-dependent magnetic field \cite{Saito}, and the DCE of phonons in a gas of laser-cooled atoms with time-dependent effective charge \cite{Dodonov10} have been studied.


Over the past decade, we have witnessed remarkable and rapid progress in the field of cavity quantum optomechanics and it is currently subject to intense research investigations (for a recent review, see, e.g., \cite{Aspelmeyer}). 
Optomechanical systems in which the electromagnetic radiation pressure is linearly or quadratically coupled to a mechanical oscillator (MO) have been widely employed in a large variety of applications; for example displacement and force sensing \cite{xsensing1,CQNCPRL,CQNCPRX,CQNCmeystre,CQNCmaximilian,aliNJP,complexCQNC}, ground-state cooling of the vibrational modes of an MO \cite{ground state cooling, Sideband cooling,Laser cooling,Teufel,Chan}, generation of entanglement \cite{Palomaki2,Paternostro,genes-entangelment}, synchronization of MOs \cite{Mari1,MianZhang,Bagheri,Shlomi,grebogi,foroud} and generation of nonclassical states of the mechanical and optical modes \cite{Borkje,Hammerer}. Furthermore, hybrid optomechanical systems consisting of atomic BECs have attracted considerable attention on studies such as nonlinear effects of atomic collisions on the optomechanical properties of a BEC trapped inside an optical cavity \cite{dalafi1}, control of optical bistability, cooling and entanglement using nonlinear atom-atom interaction \cite{dalafi2}, effects of the phase noise of the driving laser on the bipartire entanglement in a BEC-hybridized optomechanical cavity  \cite{dalafi3}, and also the effects of intrinsic cross-Kerr nonlinearity of an atomic BEC inside an optical cavity on the steady-state behavior of the system \cite{dalafi4}.

Nevertheless to our knowledge, there are only few recent investigations regarding the occurrence of DCE in optomechanical systems. Among them, we can mention Ref. \cite{Salvatore} as an MIDCE proposal in which it has been shown that a considerable number of photons  can be generated out of the vacuum for mechanical frequencies equal to or lower than the cavity-mode frequencies in the strong coupling regime without any time-dependent modulation. However, it is hardly achievable in the current technologies and it seems still to be far fetch. 
Furthermore, two theoretical schemes have been proposed to realize the PDCE in cavity optomechanical systems in which the quantum vacuum is parametrically amplified by using external modulation techniques: the scheme of realizing the DCE in a non-stationary quantum well-assisted hybrid optomechanical cavity driven by an amplitude-modulated external laser pump \cite{Mahajan}, and the scheme for the realization of the DCE of phonon excitation in the so-called membrane-in-the-middle optomechanical system \cite{Thompson} which is pumped by a far-detuned driving classical laser and also is precooled down to almost the ground state by an auxiliary cavity mode \cite{motazedi2}. 

Motivated by the above-mentioned interesting features of hybrid optomechanical systems in the field of DCE, in this paper we propose a theoretical scheme for controllable generation of the Casimir photons and phonons in an optomechanical cavity with a moving end mirror containing an interacting cigar-shaped BEC. Here, to achieve parametric amplification of quantum vacuum fluctuations, we consider an external time modulation on the spring coefficient of the MO which will be shown to be directly responsible for the amplification of the mechanical vacuum fluctuations of the MO, and indirectly responsible for the amplification of the atomic vacuum fluctuations of the BEC. In addition, by considering the time modulation of the two-body atomic collisions frequency we will show that it not only causes directly the amplification of the atomic vacuum fluctuations, but also indirectly induces the amplification of the mechanical vacuum fluctuations of the MO. 
These two types of modulations can lead to the controllable generation of the mechanical- and Bogoliubov-type Casimir phonons in the steady-state.
 Moreover, in the regime of largely different optomechanical coupling strengths of the cavity field to the BEC and the MO corresponding to the \textit{coherent} regime a considerable number of Casimir photons can be generated due to the induced amplification of quantum vacuum fluctuations when the cavity is driven by a red-detuned laser. Furthermore, by analyzing the quantum Langevin equations (QLEs) in the Fourier space, we identify an induced frequency-dependent amplification coefficient which is analogous to the frequency-dependent amplitude in an optical parametric amplifier (OPA) and also is controllable via the parameters of the external modulations as well as cooprativities.

The paper is organized as follows. In Sec. \ref{sec2}, we describe the physical model of the system under consideration and illustrate the modulation scenarios together with their corresponding parametric amplification-type Hamiltonians. 
In Sec. \ref{sec3}, the QLEs are derived and linearized around the semiclassical steady state, and in Sec. \ref{sec4} by solving the QLEs in the frequency space we find the self-energies and induced parametric amplifications of the photonic and phononic quantum vacuum fluctuations. In Sec. \ref{sec5}, we determined the steady-state mean number of generated Casimir photons and phonons in different regimes by solving the Lyapunov equation numerically. Finally, our conclusions and outlooks are summarized in Sec. \ref{sec6}.

\section{THEORETICAL DESCRIPTION OF THE SYSTEM}\label{sec2}

\begin{figure}
	\includegraphics[width=8.5cm]{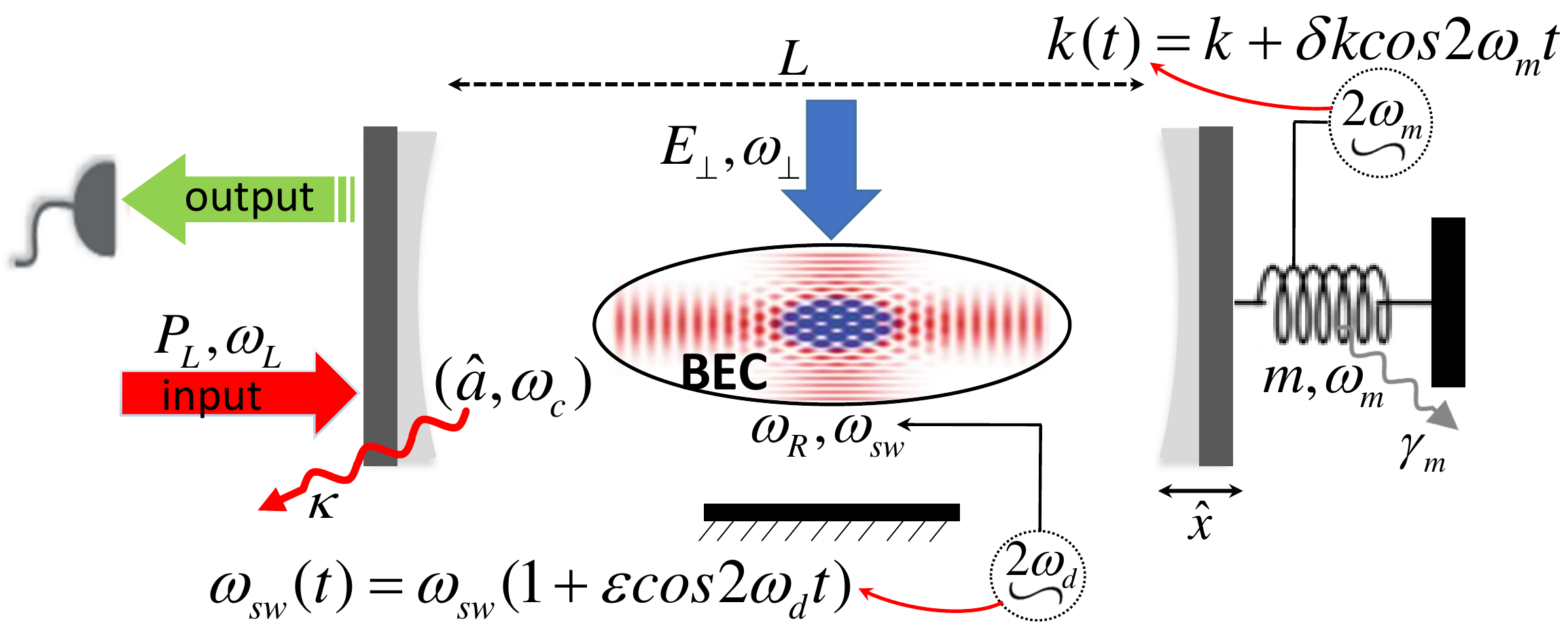}
	\caption{(Color online) Schematic diagram of a BEC-hybridized driven optomechanical cavity with a movable end mirror whose spring coefficient, $ k(t) $, is modulated parametrically. It is also assumed that the two-body \textit{s}-wave scattering frequency of atomic collision, $ \omega_{sw}(t) $, is parametrically modulated. Both the moving mirror and the BEC interact with the single mode of the cavity in the red-detuned regime, although there is no direct interaction between the BEC and the mirror. The coherent modulations lead to the generation of Casimir photons and phonons which are controllable via the optomechanical and modulation parameters.}
	\label{fig1}
\end{figure}
As shown in Fig. (\ref{fig1}), we consider a driven \textit{hybrid} optomechanical cavity with lenghth $ L $ whose end mirror is free to oscillate at mechanical frequency  $ \omega_m $ .The cavity contains a cigar-shaped BEC of $ N $ ultracold two-level atoms with mass $ m_a $ and transition frequency $ \omega_a $. The cavity is driven at rate $E_L= \sqrt{\kappa P_L / \hbar \omega_L}$ through the fixed mirror by a laser with frequency $\omega_{L}$ and wavenumber $k_{0}=\omega_{L}/c$ ($P_{L}$ is the laser power and $\kappa$ is the cavity decay rate). The total many-body Hamiltonian of the system is given by


\begin{eqnarray} \label{H1total}
&& \hat H= \hat H_{cav}+ \hat H_{m}+ \hat H_{opt} + \hat H_{mod} + \hat H_{BEC}.
\end{eqnarray}
The first three terms in Eq. (\ref{H1total}) describe, respectively, the driven cavity field, the free mechanical resonator, and the optomechanical interaction which in the frame rotating at the driving laser frequency $ \omega_L $ can be written as
\begin{subequations}\label{H_c,m,opt}
\begin{eqnarray}
&& \hat H_{cav}=\hbar \Delta_c \hat a^\dag \hat a + i\hbar E_L (\hat a^\dag - \hat a), \\
&& \hat H_m=\frac{\hat p^2}{2m} + \frac{1}{2} m \omega_m^2 x^2 =\hbar \omega_m \hat b^\dag \hat b, \\
&& \hat H_{opt}= -\hbar g_0 \hat a^\dag \hat a (\hat b + \hat b^\dag),
\end{eqnarray}
\end{subequations}
where $ \hat a $ ($ \hat b $) is the annihilation operator of the cavity (mechanical) mode and $ \Delta_c= \omega_c-\omega_L $ is the detuning of the cavity mode from the driving laser frequency. Here, we have modeled the mirror as a single-mode MO of intrinsic frequency $ \omega_m $, effective mass $ m $, and damping rate $ \gamma_m $. The canonical position and momentum of the MO are $ \hat x= x_{zp}(\hat b + \hat b^\dag) $ and $ \hat p= \hbar (\hat b - \hat b^\dag)/2ix_{zp} $, respectively, with $ x_{zp}= \sqrt{\hbar / 2m\omega_m} $ being the zero-point position fluctuation. $ g_0=x_{zp} \omega_c /L $ stands for the single-photon optomechanical coupling.
The single-mode approximation for the mechanical and optical fields is an appropriate simplification provided that the cavity free spectral range is much larger than the mechanical frequency \cite{singlecavitymode} and the detection bandwidth is chosen such that it includes only a single isolated mechanical resonance and mode-mode coupling is negligible \cite{singlemechanicalmode}. The last two terms in Eq. (\ref{H1total}) describe, respectively, the time modulation of the spring coefficient of the MO and the Hamiltonian of the atomic BEC whose explicit forms will be determined in the following.


\subsection{Modulation of the mechanical spring coefficient}
We assume that the MO is parametrically driven by modulating its spring coefficient at twice its natural frequency, i.e., $k(t)= k+\delta k \cos(2\omega_{m} t +\varphi_m) $ ($ \varphi_m $ being the phase of external modulation) which is equivalent to the modulation of the mechanical frequency [ see Fig. (\ref{fig1})]. In this way, the Hamiltonian of the MO can be written as \cite{optomechanics with two phonon driving}
\begin{eqnarray} \label{H_m(t)}
&& \hat H_m(t)=\frac{ \hat p^2}{2m} + \frac{1}{2} k(t) x^2=\hat H_m + \hat H_{mod},
\end{eqnarray}
where 
\begin{eqnarray} \label{H_mod(t)}
&& \hat H_{mod}(t)= \frac{1}{2} \delta k \cos(2\omega_m t) x^2.
\end{eqnarray}
Now, using the rotating wave approximation (RWA) over time scales longer than $ \omega_m^{-1} $ to ignore the fast rotating term in the shifted frequency of the MO, i.e., the term proportional to $\cos(2\omega_{m} t) \hat b^\dag \hat b $, the Hamiltonian of Eq. (\ref{H_mod(t)}) can be simplified as 
\begin{eqnarray} \label{H_mod}
&& \hat H_{mod}(t)= \frac{i \hbar}{2} (\lambda_m \hat b^{\dag 2}  e^{-2i\omega_m t}- \lambda_m^\ast \hat b^2 e^{2i\omega_m t}),
\end{eqnarray}
where $ \lambda_m= \vert \lambda_m \vert e^{i\varphi_m} $ with $ \vert \lambda_m \vert= \delta k x_{zp}^2 / 2\hbar $. 
The Hamiltonian of Eq. (\ref{H_mod}) describes the mechanical phonon analog of the degenerate parametric amplification (DPA) where the vibrational fluctuation of the MO plays the role of the signal mode in the DPA. Accordingly, it is expected that the coherent modulation of the mechanical spring coefficient or mechanical frequency leads to the amplification of quantum vacuum fluctuations of the MO, i.e., dynamical Casimir emission of mechanical phonons (mechanical-type Casimir phonons). Note that by fixing the phase of modulation $ \varphi_m $, it is always possible to take $ \lambda_m $ as a real number.

\subsection{Second quantized Hamiltonian of the BEC}
We now assume $ N $ ultracold two-level atoms of the BEC to be confined in a cylindrically symmetric optical trap with a transverse trapping frequency $ \omega_\bot $ and negligible longitudinal confinement frequency $ \omega_\parallel $ along the axis of the cavity (x axis). Therefore, we can describe the dynamics within an effective one-dimensional model by quantizing the atomic motional degree of freedom along the $ x $ axis only.

In the dispersive regime of  atom-field interaction where the laser pump is far-detuned from the atomic resonance ($ \Delta_a=\omega_a - \omega_L \gg \Gamma_a $ where $ \Gamma_a$ is the atomic linewidth), the excited electronic state of the atoms can be adiabatically eliminated and spontaneous emission can be neglected \cite{Ritsch}. The many-body Hamiltonian of the BEC can be written as
\begin{eqnarray} \label{H1_BEC}
&& \hat  H_{BEC}=\int_{-L/2}^{L/2}dx \hat \psi^\dag(x) \hat H_0 \hat \psi(x) +\hat H_s ,
\end{eqnarray}
in which
\begin{subequations} \label{H_0&s}
   \begin{eqnarray} \label{H_0}
     && \hat H_0= \frac{-\hbar^2}{2m_a} \frac{d^2}{dx^2}+ \hbar U_0 \cos^2(k_0 x) \hat a^\dag \hat a + V_{ext}^{\lVert}(x), \\
     && \hat H_s= \frac{1}{2} U_s  \int_{-L/2}^{L/2} dx \hat \psi^{\dag}(x)  \hat \psi^{\dag}(x) \hat \psi(x)  \hat \psi(x)  ,
   \end{eqnarray}
\end{subequations}
where $ \hat \psi(x) $ is the annihilation operator of the atomic field, $ U_0=-g_a^2/\Delta_a $ is the optical lattice barrier height per photon which represents the atomic backaction on the field in the dispersive regime, $ g_a $ is the vacuum Rabi frequency or atom-field coupling constant, $ U_s=4\pi \hbar^2 a_s/m_a $, and $ a_s $ is the two-body \textit{s}-wave scattering length	 \cite{Ritsch,Domokos}. 

In Eq. \ref{H_0&s}(a), the longitudinal trapping potential $ V_{ext}^{\lVert}(x)= m_a \omega_{\lVert}^2 x^2/2 $ can be approximately ignored. In an effective one-dimensional model this approximation is valid as long as $ \omega_\bot \gg \omega_\lVert $ so that the periodic potential of the optical lattice in the longitudinal direction is only slightly modified by $ V_{ext}^{\lVert}(x) $. Additionally, the energy arising from the atom-atom interaction has to be smaller than the energy splitting of the transverse vibrational states $ \hbar \omega_\bot $ which implies that the linear density of the condensate is smaller than $ 1/2a_s $ \cite{Morsch,Cigar}.

In the weakly interacting regime, i.e., $ U_0 \langle \hat a^\dag \hat a  \rangle \le 10\omega_R $ where $ \omega_R=\hbar k_0^2/2m_a $ is the recoil frequency of the condensate atoms, one can restrict the atomic field operator $\hat \psi(x) $ to the first two symmetric momentum side modes with momenta $ \pm 2\hbar k_0 $ which are excited by the atom-light interaction \cite{Domokos2}.  In this manner, because of the parity conservation and considering the Bogoliubov approximation \cite{Nagy}, the atomic field operator can be expanded as the following single-mode quantum field:
\begin{eqnarray} \label{si1}
&& \hat \psi(x)= \sqrt{N/L}+ \phi_2(x) \hat d ,
\end{eqnarray}
where $ \phi_2(x)= \sqrt{2/L}\cos(2k_0x) $ and the Bogoliubov mode $ \hat d $ corresponds to the quantum fluctuations of the atomic field around the classical condensate mode $ \sqrt{N/L} $. In this expansion we have not only neglected terms proportional to $ \cos(2mk_0x) $ with $ m \ge 2 $ but also the terms with nonzero quasimomenta \cite{dalafi1}. 
Based on the numerical results obtained from \cite{Cigar} even the transition probability to the state with $ m=2 $ is very low. As has been shown in Refs. \cite{MeystreBEC, dalafi qpt}, the scattering to these extra modes due to the atom-atom interaction, or interaction with  $ V_{ext}^{\lVert}(x) $, can be simulated as a damping process and may be incorporated into the noise that affects the matter field.

In the case that the system does not have parity symmetry, for example, when the BEC is trapped inside a ring cavity, one should also consider $ \rm sine $ modes which in our model have been set aside \cite{MeystreBEC2,MeystreBEC3}. By substituting the atomic field operator, Eq. (\ref{si1}), into the Hamiltonian of Eq. (\ref{H1_BEC}), we arrive at the following form for the Hamiltonian of the atomic BEC subsystem
	\begin{eqnarray}   
	&&  \hat H_{BEC}\! = \! \!  \hbar\delta_0 \hat a^\dag \hat a \!+\!  \hbar \omega_d \hat d^\dag \hat d \!+ \! \hbar G_0\hat a^\dag \hat a (\hat d  + \hat d^\dag) \! +  \! \hat H_{sw} \!\! + \hat H_{CK} , \label{H_BEC} \\
	&& \hat H_{sw}= \hbar\frac{\omega_{sw}}{4} (\hat d^2 + \hat d^{\dag 2}) , \label{H_sw} \\
	&& \hat H_{CK}=  \hbar \eta \hat a^\dag \hat a \hat d^\dag \hat d  , \label{H_ck}
	\end{eqnarray}	
where $ \delta_0=NU_0/2 $, $ \omega_d=4\omega_R + \omega_{sw} $ is the effective frequency of the Bogoliubov mode in the atomic BEC, $ G_0=\sqrt{2N}U_0/4 $ is the strength of an optomechanical-like coupling between the Bogoliubov mode of the BEC and the intracavity field, $ \omega_{sw}= 8\pi\hbar N a_s/(m_a L w^2) $ is the \textit{s}-wave scattering frequency of atom-atom interaction ( with $ w $ being the waist radius of the optical mode), and  $ \eta=U_0/2 $ is the cross-Kerr (CK) coefficient.

The first term in Eq. (\ref{H_BEC}) leads to a shift of the cavity detuning in Eq. \ref{H_c,m,opt}(a), i.e., $ \Delta_c \to \Delta_0=\Delta_c + NU_0/2 $ which can be interpreted as an effective Stark-shifted detuning. The second term describes the energy of the Bogoliubov mode $ \hat d $. The third term is an optomechanical-like interaction which corresponds to the linear radiation pressure coupling of the Bogoliubov mode and the optical field. In this manner, the Bogoliubov mode plays the role of another MO. The fourth term is the atom-atom interaction energy  which plays the role of the \textit{atomic parametric amplifier} and is responsible for the generation of atomic squeezed state. The last term denotes the CK nonlinear coupling between the intracavity field and the Bogoliubov mode. As has been shown in Ref. \cite{dalafi4}, in the case of $ \eta/ G_0\ll 1 $ the effect of this term is negligible. Therefore, we will ignore the CK term in our calculations.

\subsection{Modulation of the atomic collisions}
Here, we consider a time modulation of the atomic collisions at twice of the Bogoliubov mode of the BEC, i.e., we assume $ \omega_{sw}(t)= \omega_{sw}\big(1+\varepsilon \cos (2\omega_d t+\varphi_d)\big ) $ where $ \varepsilon $ and $ \varphi_d $ denote, respectively, the amplitude and the phase of modulation. It can be realized by the time modulation of the scattering length $ a_s $ via the modulation of the electromagnetic trap, or the modulation of the density of the BEC by changing the trap stiffness via the intensity modulation of the pump laser, thereby modulating the speed of sound \cite{Jaskula}.
Applying the RWA on the Hamiltonian (\ref{H_sw}) one can find 
\begin{eqnarray} \label{H_sw_mod}
&& \hat H_{sw}(t) \approx \frac{i \hbar}{2} (\lambda_d \hat d^{\dag 2}  e^{-2i\omega_d t}- \lambda_d^\ast \hat d^2 e^{2i\omega_d t}),
\end{eqnarray}
where $ \lambda_{d}=-i\varepsilon \omega_{sw} e^{-i\varphi_d}/4 $ can be taken real by fixing the phase $ \varphi_d $ . As can be seen, the Hamiltonian (\ref{H_sw_mod}) describes a Bogoliubov phonon analog of the DPA where the quantum fluctuation of the Bogoliubov mode plays the role of the signal mode in the DPA. It means that the coherent modulation of the frequency of the atomic collisions leads to the amplification of quantum vacuum fluctuations of the Bogoliubov mode, i.e., the dynamical Casimir emission of Bogoliubov phonons (Bogoliubov-type Casimir phonons).

\section{system dynamics}\label{sec3}
Based on the above considerations, one can rewrite the total Hamiltonian of the system in the frame rotating at the driving laser frequency ($ \omega_L $) as follows:
\begin{eqnarray}
&& \mathcal{\hat H}_{tot}=\hbar \Delta_0 \hat a^\dag \hat a + \hbar \omega_m \hat b^\dag \hat b + \hbar \omega_d \hat d^\dag \hat d + i \hbar E_L(\hat a^\dag - \hat a ) \nonumber \\
&& \qquad \qquad \qquad -\hbar g_0 \hat a^\dag \hat a (\hat b+\hat b^\dag) + \hbar G_{0} \hat a^\dag \hat a (\hat d +\hat d^\dag)  \nonumber \\
&& \qquad \qquad \qquad + i \frac{\hbar }{2} (\lambda_m\hat b^{\dag2} e^{-2i\omega_m t}+ \lambda_m^\ast \hat b^2 e^{2i\omega_m t}) \nonumber \\
&& \qquad \qquad \qquad + i  \frac{ \hbar}{2} (\lambda_d \hat d^{\dag 2}  e^{-2i\omega_d t}- \lambda_d^\ast \hat d^2 e^{2i\omega_d t}) .
\end{eqnarray}
The corresponding QLEs can be written as
\begin{subequations} \label{Langevin1}
\begin{eqnarray}
&&\!\!\!\!\!\!\!\!\!\!\!\!\!\! \dot {\hat a} \! =\!\! -(i \Delta_0 \!+\! \frac{\kappa}{2}) \hat a\! + \! i g_0 (\hat b \!+\! \hat b^\dag) \hat a\!  - \! i G_0 (\hat d \! + \!\hat d^\dag) \hat a \!+\! E_L \!\!+ \!  \! \sqrt{\kappa} \hat a_{in} , \\
&&\!\!\!\!\!\!\!\!\!\!\!\!\!\!   \dot {\hat b} \! =\! -(i\omega_m +\frac{\gamma_m}{2}) \hat b +ig_0 \hat a^\dag \hat a \!+\! \lambda_m e^{-2i\omega_m t} \hat b^\dag \! + \! \sqrt{\gamma_m} \hat b_{in}, \\
&&\!\!\!\!\!\!\!\!\!\!\!\!\!\!   \dot {\hat d} = -(i \omega_d + \dfrac{\gamma_d}{2}) \hat d -i G_0 \hat a^\dag \hat a + \lambda_d e^{-2i\omega_d t} \hat d^\dag \! + \!\sqrt{\gamma_d} \hat d_{in}.
\end{eqnarray}	
\end{subequations}

Here, the cavity-field quantum vacuum fluctuation $\hat a_{in}(t)$ satisfies the Markovian correlation functions, i.e., $\langle\hat a_{in}(t)\hat a_{in}^{\dagger}(t^{\prime})\rangle=(\bar n_{ph}+1)\delta(t-t^{\prime})$ and $ \langle\hat a_{in}^{\dagger}(t)\hat a_{in}(t^{\prime})\rangle=\bar n_{ph}\delta(t-t^{\prime}) $ with the average thermal photon number $\bar n_{ph}$ which is nearly zero at optical frequencies \cite{Gardiner}. We also assume that the Brownian noises  $ \hat b_{in} $ and $ \hat d_{in} $ affecting, respectively, the mechanical mode and the Bogoliubov mode of the BEC (regarded as a formal analog of an MO) have Markovian behavior which is valid for oscillators with high quality factors \cite{MeystreBEC, vitali noise membrane}. Their correlation functions can be written as $ \langle\hat b_{in}(t)\hat b_{in}^{\dagger}(t^{\prime})\rangle=(\bar n_{m}+1)\delta(t-t^{\prime}) $, $ \langle\hat b_{in}^{\dagger}(t)\hat b_{in}(t^{\prime})\rangle=\bar n_{m}\delta(t-t^{\prime}) $,
 $ \langle\hat d_{in}(t)\hat d_{in}^{\dagger}(t^{\prime})\rangle=(\bar n_{d}+1)\delta(t-t^{\prime}) $, and $ \langle\hat d_{in}^{\dagger}(t)\hat d_{in}(t^{\prime})\rangle=\bar n_{m}\delta(t-t^{\prime}) $  where $ \bar n_m=[\exp(\hbar \omega_m/k_B T)-1]^{-1} $  and $ \bar n_d=[\exp(\hbar \omega_d/k_B T)-1]^{-1} $ are the thermal excitations of the mechanical and Bogoliubov modes, respectively. It should be noted that the noise sources are assumed to be uncorrelated for the different modes of both the matter and light fields.

The QLEs (\ref{Langevin1}a)-(\ref{Langevin1}c) can be solved analytically by adopting a linearization procedure in which the operators are expressed as the sum of their classical mean values and small fluctuations, i.e., $ \hat o = \bar o + \delta \hat o $ with $ \langle \delta \hat o^\dag \delta \hat o \rangle / \langle \hat o^\dag \hat o \rangle \ll 1  $ ($ o=a,b,d $).
The linearized QLEs describing the dynamics of the quantum fluctuations are given by
\begin{subequations} \label{Langevin-linearized}
	\begin{eqnarray}
	&&\! \! \!\! \! \!\!\!\!\!\!\! \delta \dot { \hat a}\! =\! -(i \bar \Delta_0 \!+\! \frac{\kappa}{2}) \delta \hat a \!+\! i g (\delta \hat b \! + \! \delta  \hat b^\dag) \! - \! i G (\delta \hat d \! +\! \delta \hat d^\dag) \! + \! \sqrt{\kappa} \hat a_{in} , \\
	&&\! \! \!\! \! \!\!\!\!\!\!\! \delta \dot {\hat b} \! =\! -(i\omega_m \!\! +\! \! \frac{\gamma_m}{2}) \delta \hat b \! + \!ig (\! \delta \hat a^\dag \! + \! \delta \hat a ) \!+\! \lambda_m e^{-2i\omega_m t} \! \delta \hat b^\dag \! \!+ \! \!\! \sqrt{\gamma_m} \hat b_{in}, \\
	&&\! \! \!\! \! \!\!\!\!\!\!\! \delta \dot {\hat d} \! =\! -(i \omega_d \! + \! \dfrac{\gamma_d}{2}) \delta \hat d  \! -i G(\delta \hat a^\dag \!\! + \! \delta \hat a ) \! +\!\! \lambda_d e^{-2i\omega_d t} \! \delta \hat d^\dag \! \! +\!\! \sqrt{\gamma_d} \hat d_{in},  
    \end{eqnarray}	
\end{subequations}
where $ g=g_0 \bar a $ and $ G = G_0 \bar a $ are the enhanced-optomechanical coupling strengths of the moving mirror and the Bogoliubov mode of the BEC to the intracavity field, respectively, and $ \bar \Delta_0 = \Delta_0 -2g_0 {\rm{Re}}  (\bar b) +2 G_{0} {\rm{Re}}(\bar d ) \approx \Delta_0 $.

From the linearized QLEs (\ref{Langevin-linearized}a)-(\ref{Langevin-linearized}c), one can deduce the following linearized Hamiltonian in the Schr\"{o}dinger picture
\begin{eqnarray} \label{H-Linear}
&&  \mathcal{\hat H}_L= \hbar \bar \Delta_0 \delta \hat a^\dag \delta \hat a + \hbar \omega_m \delta \hat b^\dag \delta \hat b + \hbar \omega_d \delta \hat d^\dag \delta \hat d \nonumber \\
&& \qquad  - \hbar g (\delta \hat a + \delta \hat a^\dag) (\delta \hat b + \delta \hat b^\dag)+\hbar G (\delta \hat a + \delta \hat a^\dag) (\delta \hat d + \delta \hat d^\dag) \nonumber \\
&& \qquad + i\frac{\hbar }{2}(\lambda_m \delta \hat b^{\dag 2} e^{-2i\omega_m t} - \lambda_m^\ast \delta \hat b^2 e^{2i\omega_m t}) \nonumber \\
&& \qquad + i\frac{\hbar }{2}(\lambda_d \delta \hat d^{\dag 2} e^{-2i\omega_d t} - \lambda_d^\ast \delta \hat d^2 e^{2i\omega_d t}),
\end{eqnarray}
or in the interaction picture
\begin{eqnarray} \label{H-Linear-interaction}
&&  \mathcal{\hat H}_L^{(I)}= - \hbar g (\delta \hat a e^{-i\bar \Delta_0 t} + \delta \hat a^\dag e^{i\bar \Delta_0 t}) (\delta \hat b e^{-i\omega_m t}+ \delta \hat b^\dag e^{i\omega_m t}) \nonumber  \\
&& \qquad \quad +\hbar G (\delta \hat a e^{-i\bar \Delta_0 t}  + \delta \hat a^\dag e^{i\bar \Delta_0 t} ) (\delta \hat d  e^{-i\omega_d t} + \delta \hat d^\dag e^{i\omega_d t}) \nonumber \\
&& \qquad \quad + i\frac{\hbar }{2}(\lambda_m \delta \hat b^{\dag 2} - \lambda_m^\ast \delta \hat b^2 ) + i\frac{\hbar }{2}(\lambda_d \delta \hat d^{\dag 2} - \lambda_d^\ast \delta \hat d^2).
\end{eqnarray}

To proceed further, we assume that the frequency of the Bogoliubov mode of the BEC, $ \omega_d $, can be matched to the mechanical frequency ($ \omega_d \approx \omega_m $) via the atomic collisions frequency $ \omega_{sw} $ or through the recoil frequency $ \omega_R $ via the driving laser frequency. Therefore, we restrict ourselves to the red-detuned regime of cavity optomechanics where $ \Delta_0 \approx \omega_m \approx \omega_d $. In this regime and within the RWA where the BEC-cavity mode as well as the MO-cavity mode couplings are analogous to the beam-splitter interaction, Eqs. (\ref{Langevin-linearized}a)-(\ref{Langevin-linearized}c) can be rewritten as follows
\begin{subequations} \label{Langevin-BS}
	\begin{eqnarray}
	&& \delta \dot { \hat a} = -\frac{\kappa}{2} \delta \hat a + i g  \delta \hat b - i G \delta \hat d + \sqrt{\kappa} \hat a_{in} , \\
	&& \delta \dot {\hat b}  = -\frac{\gamma_m}{2} \delta \hat b + i g \delta \hat a + \lambda_m \delta \hat b^\dag +  \sqrt{\gamma_m} \hat b_{in}, \\
	&& \delta \dot {\hat d}  = - \dfrac{\gamma_d}{2} \delta \hat d  -i G  \delta \hat a + \lambda_d \delta \hat d^\dag  +  \sqrt{\gamma_d} \hat d_{in}.
	\end{eqnarray}
\end{subequations}
Now, by defining the quadratures $ \delta \hat X_o=(\hat o+\hat o^{\dagger})/\sqrt{2} $ and $ \delta \hat P_o=(\hat o-\hat o^{\dagger})/\sqrt{2}i $ ($ o=a,b,d $) we can express the linearized Eqs. (\ref{Langevin-BS}a)-(\ref{Langevin-BS}c) in the compact matrix form as
\begin{eqnarray} \label{udot}
&&\delta \dot {\hat u}(t)= A ~\delta \hat u(t) + \hat N(t),
\end{eqnarray}
where the vector of continuous-variable fluctuation operators and the corresponding vector of noises are given by $ \hat u=(\delta \hat X_a,\delta \hat P_a,\delta \hat X_b, \delta \hat P_b,\delta \hat X_d,\delta \hat P_d)^T $ and $ \hat N=(\sqrt{\kappa}\hat X_a^{in},\sqrt{\kappa}\hat P_a^{in},\sqrt{\gamma_m}\hat X_b^{in},\sqrt{\gamma_m}\hat P_b^{in}, \sqrt{\gamma_d} \hat X_d^{in},\sqrt{\gamma_d} \hat P_d^{in} )^T $, respectively. Here, $ \hat X_o^{in}=(\hat o_{in}+\hat o_{in}^{\dagger})/\sqrt{2} $ and $ \hat P_o^{in}=(\hat o_{in}-\hat o_{in}^{\dagger})/\sqrt{2}i $ ($ o=a,b,d $).
Furthermore, the time-independent drift matrix $ A $ is given by 
\begin{eqnarray} \label{A}
&&\!\!\!\!\!\!\!\!\!\!\!\!  A\!=\! \left( \begin{matrix}
{-\frac{\kappa}{2}} & {0} & {0} & {-g} & {0} & {G}  \\
{0} & {-\frac{\kappa}{2}} & {g} & {0} & {-G} & {0}  \\
{0} & {-g} & {\!\lambda_m\!-\!\frac{\gamma_m}{2}} & {0} & {0} & {0}  \\
{g} & {0} & {0} & {-(\lambda_m\!+\!\frac{\gamma_m}{2})} & {0} & {0}  \\
{0} & {G} & {0} & {0} & {\lambda_d -\frac{\gamma_d}{2}} & {0}  \\
{-G} & {0} & {0} & {0} & {0} & {-(\lambda_d\!+\!\frac{\gamma_d}{2} )}  \\
\end{matrix} \right).
\end{eqnarray}
In the steady state ($ \kappa t \gg 1 $) with the condition $ \kappa \gg \gamma_{m,d} $, the RWA leads to the good-cavity limit, i.e., $ \omega_m \gg \kappa $. The formal solution of Eq. (\ref{udot}) can be obtained by integration. On the other hand, the symmetric covariance matrix with entries given by $ V_{ij}=\langle \hat u_i(t) \hat u_j(t) +\hat u_j(t) \hat u_i(t) \rangle /2$ fully characterizes the quantum correlations of the system quadratures whose dynamics can be expressed as the following equation
\begin{eqnarray} \label{Vdot}
&& \frac{d V}{dt}=  A ~ V + V ~ A^T + D , 
\end{eqnarray}
where $ D $ is the diffusion matrix defined as
\begin{eqnarray} \label{D}
&& D_{ij} \delta (t-t')=\frac{1}{2} \langle \hat N_i(t) \hat N_j(t')+ N_j(t') N_i(t) \rangle .
\end{eqnarray}
 In the steady state the covariance matrix $ V $ solves the Lyapunov equation
  \begin{eqnarray} \label{V}
  &&   A ~ V + V ~ A^T = - D, 
  \end{eqnarray}
with
\begin{eqnarray} 
&& \! \!\!\!\!\!\!\!\!\!\!\!\!\!\! D \!= \! {\rm {Diag}} \Big [\frac{\kappa}{2},\frac{\kappa}{2},\frac{\gamma_m}{2}n^{\prime}_m,\frac{\gamma_m}{2}n^{\prime}_{m},\frac{\gamma_d}{2}n^{\prime}_d, \frac{\gamma_d}{2}n^{\prime}_d\Big], \!\!
\label{D_ij_standard}
\end{eqnarray}
where $ n^{\prime}_m=2\bar n_m+ 1 $ and  $ n^{\prime}_d=2\bar n_d +1 $.
 
\section{QLE{\scriptsize s} in Fourier space}\label{sec4}
The set of differential equations (\ref{Langevin-BS}) can be solved in the frequency space by the Fourier transformation, $ \hat O(t)= (1/\sqrt{2\pi}) \int_{-\infty}^{\infty} d\omega \hat O(\omega) e^{-i\omega t} $. The solution can be written in the following form
\begin{subequations}
\begin{eqnarray}
&& \!\! -i\omega \delta \hat a(\omega)  =\! -[\kappa/2 \! +\! i\Sigma_a(\omega)] \delta \hat a(\omega) \! + \!  \tilde \lambda_a(\omega) \delta \hat a^\dagger(\omega) \! + \!\!\! \sqrt{\kappa} \hat A_{in}(\omega) , \label{a_omega} \nonumber  \\
&& \\
&&\!\! -i\omega \delta \hat b(\omega)=\! -[\gamma_m/2 \!- \! i\Sigma_b(\omega)] \delta \hat b(\omega) \! + \! \tilde \lambda_b(\omega) \delta \hat b^\dagger(\omega) \! + \!\!\! \sqrt{\gamma_m} \hat B_{in}(\omega), \label{b_omega} \nonumber  \\
&& \\
&& \!\! -i\omega \delta \hat d(\omega)= \! -[\gamma_d/2 \! + \! i \Sigma_d(\omega)] \delta \hat d(\omega) \! + \! \tilde \lambda_d(\omega) \hat d^\dagger(\omega)\! + \!\!\! \sqrt{\gamma_d} \hat D_{in}(\omega). \label{d_omega}  \nonumber \\ 
\end{eqnarray}
\end{subequations}
Here, the self-energies are given by
\begin{subequations}
	\begin{eqnarray}
	&& i \Sigma_a(\omega)= g^2 \frac{\gamma_m/2-i \omega}{(\gamma_m/2-i \omega)^2-\vert \lambda_m \vert^2} + G^2 \frac{\gamma_d/2-i \omega}{(\gamma_d/2-i \omega)^2-\vert \lambda_d \vert^2} , \nonumber \\
	&& \\
	&& i\Sigma_b(\omega)= g^2 \frac{\kappa/2 -i(\omega-\Sigma_s(\omega))}{[\kappa/2-i(\omega-\Sigma_s(\omega))]^2- \bar \Lambda_s^2(\omega)} , \label{S_b} \\
	&&  i\Sigma_d(\omega)= G^2 \frac{\kappa/2 - i(\omega-\Sigma_m(\omega)) }{[\kappa/2 - i(\omega-\Sigma_m(\omega))]^2 - \Lambda_m^2(\omega)} ,
	\end{eqnarray}
\end{subequations}
with
\begin{subequations}
\begin{eqnarray} 
&& i \Sigma_s(\omega)=G^2\frac{\gamma_d/2- i\omega}{(\gamma_d/2-i\omega)^2 - \vert \lambda_d \vert^2} ~, \label{sigma_s}  \\
&& i\Sigma_m(\omega)= g^2 \frac{\gamma_m/2-i\omega}{(\gamma_m/2-i\omega)^2 - \vert \lambda_m \vert^2} ~ , \label{sigma_m} \\
&& \bar \Lambda_s(\omega)=\lambda_d G^2 \frac{1}{(\gamma_d/2-i\omega)^2 - \vert \lambda_d \vert^2} ~, \label{lambda_s} \\
&& \Lambda_m(\omega)=\lambda_m  g^2 \frac{1}{(\gamma_m/2-i\omega)^2 - \vert \lambda_m \vert^2} \label{lambda_m} ~.
\end{eqnarray}
\end{subequations}
The \textit{induced} frequency-dependent parametric amplification factors which are analogous to the gain factors in the conventional DPA are given by 
\begin{subequations}
	\begin{eqnarray}
     &&\!\!\!\!\!\!\!\!\! \tilde \lambda_a(\omega)=\frac{ g^2  \lambda_m}{(\gamma_m/2-i \omega)^2-\vert \lambda_m \vert^2} + \frac{G^2 \lambda_d}{(\gamma_d/2-i \omega)^2-\vert \lambda_d \vert^2} , \label{lambda_a_omega} \\
     &&\!\!\!\!\!\!\!\!\! \tilde \lambda_b(\omega)= \lambda_m + g^2 \frac{\bar \Lambda_s(\omega)}{[\kappa/2-i(\omega-\Sigma_s(\omega))]^2- \bar \Lambda_s^2(\omega)} ~, \label{lambda_b_omega} \\
     &&\!\!\!\!\!\!\!\!\! \tilde \lambda_d(\omega)= \lambda_{d} + G^2 \frac{\Lambda_m(\omega)}{[\kappa/2 - i(\omega-\Sigma_m(\omega))]^2 - \Lambda_m^2(\omega)} ~,  \label{lambda_d_omega}
	\end{eqnarray}
\end{subequations}
and are responsible for the generation of the Casimir photons or phonons. 
Moreover, the generalized noise operators are given by 
\begin{subequations}
\begin{eqnarray}
&& \hat A_{in}= \hat a_{in} + ig \sqrt{\frac{\gamma_m}{\kappa}} \frac{\gamma_m/2-i \omega}{(\gamma_m/2-i \omega)^2-\vert \lambda_m \vert^2} \left[ \hat b_{in} + \frac{\lambda_m}{\gamma_m/2-i\omega} \hat b^\dagger_{in}  \right] \nonumber \\
&& \qquad -iG \sqrt{\frac{\gamma_d}{\kappa}} \frac{\gamma_d/2-i \omega}{(\gamma_d/2-i \omega)^2-\vert \lambda_d \vert^2} \left[ \hat d_{in} + \frac{\lambda_s}{\gamma_d/2-i\omega} \hat d^\dagger_{in}  \right] , \\
&& \hat B_{in}(\omega)= \hat b_{in}(\omega) +\frac{1}{g} \sqrt{\frac{\kappa}{\gamma_m}} \left[ i\bar \Lambda_b(\omega) \hat a_{in}^\dag (\omega) - \Sigma_b(\omega) \hat a_{in}(\omega) \right] \nonumber  \\ 
&&\quad -\frac{i}{gG} \sqrt{\frac{\gamma_d}{\gamma_m}} \Sigma_s(\omega) \left[ \left(\frac{\lambda_d}{\gamma_d/2 - i\omega} \bar \Lambda_b(\omega) - i\Sigma_b(\omega) \right) \hat d_{in}(\omega)  \right. \nonumber \\
&&\qquad\quad\qquad\quad  \left.  \left( \bar \Lambda_b(\omega)- i\Sigma_b(\omega) \frac{\lambda_d}{\gamma_d/2 - i\omega}  \right) \hat d_{in}^\dag(\omega) \right], \\
&& \hat D_{in}(\omega)= \hat d_{in}(\omega) + \sqrt{\frac{\kappa}{\gamma_d}}  \left[ \frac{\Sigma_d(\omega)}{G} \hat a_{in}(\omega) - i G \bar \Lambda_d(\omega) \hat a_{in}^\dag(\omega) \right] \nonumber \\
&& \quad -\frac{\Sigma_m(\omega)}{g}(\omega)  \sqrt{\frac{\gamma_m}{\gamma_d}} \left[ \frac{\Sigma_d(\omega)}{G} \hat b_{in}(\omega) + i G \bar \Lambda_d(\omega)  \hat b_{in}^\dagger(\omega)   \right],
\end{eqnarray}
\end{subequations}
where  $  \bar \Lambda_b(\omega)= (\tilde \lambda_b(\omega)-\lambda_m) /g^2 $ and $ \bar \Lambda_d(\omega)=(\tilde \lambda_d(\omega)-\lambda_d) /G^2 $. 


 As is evident, by turning on the modulation (i.e., $\lambda_{m,d} \neq 0$), the mechanics and the atomic collisions do indeed mediate a parametric-amplifier-like effective squeezing interaction for the cavity mode which is also frequency-dependent (see Eq. (\ref{lambda_a_omega})), unlike the case of a conventional DPA. More importantly, the time  modulation of the atomic collisions and parametric driving of the MO simultaneously amplify the vacuum fluctuations of the cavity field resulting in the generation of the Casimir photons in the \textit{coheren}t regime, which will be detailed in Sec. \ref{sec5}. Furthermore, as we will show, the time modulation of the atomic collisions (mechanical mode) amplifies indirectly the mechanical (atomic) vacuum fluctuations through the second term in Eq. (\ref{lambda_b_omega}) (Eq. (\ref{lambda_d_omega})) which leads to the dynamical Casimir emission of the mechanical (Bogoliubov) phonons.

Let us now find the effective damping rates of the phononic modes of the moving mirror and the BEC. We are interested in the on-resonance case, $ \omega= 0$, where the shifted mechanical frequency is zero and $ \Gamma_{op}^{m(d)}=-2{\rm Im} \Sigma_{b(d)}(\omega) $ is maximum. In the on-resonance case the effective mechanical (atomic) damping rate $ \Gamma_{m(d)} $ can be given by 
\begin{eqnarray} \label{gamma_eff}
&& \Gamma_{m(d)}= \gamma_{m(d)}+\Gamma_{op}^{m(d)}=\gamma_{m(d)} [1+\mathcal{C}_{m(d)}] ~ ,
\end{eqnarray}
where $ \mathcal{C}_m $($ \mathcal{C}_d $) is the collective optomechanical cooperativity associated with the moving mirror (Bogoliubov mode),
\begin{eqnarray}
&&  \mathcal{C}_{m(d)}=  \mathcal{C}_{0(1)} \frac{1+ \mathcal{C}_{1(0)} - \xi_{d(m)}^2}{(1+ \mathcal{C}_{1(0)}-\xi_{d(m)}^2)^2- \xi_{d(m)}^2  \mathcal{C}_{1(0)}^2} ~,
\end{eqnarray}
where $\mathcal{C}_0=4g^2/\kappa \gamma_m$ and $\mathcal{C}_1= 4G^2/\kappa \gamma_d $ are the optomechanical and opto-atomic cooperativities, respectively, and $ \xi_{d(m)}=2\lambda_{d(m)}/\gamma_{d(m)}$. 
Evidently, if $ \xi_{d(m)} \to 0 $ then $ \mathcal{C}_{m(d)} \to \mathcal{C}_{0(1)}/(1+\mathcal{C}_{1(0)})$. Both collective cooperativities $ \mathcal{C}_{m} $ and $ \mathcal{C}_{d} $ can be controlled by the time modulations of the MO and/or the BEC as well as through the individual cooperativities $ \mathcal{C}_{0} $ and $ \mathcal{C}_{1} $.

Based on  the Routh-Hurwitz criterion \cite{Routh} for the optomechanical stability condition, the parameters $ \lambda_{m} $ and $ \lambda_{d} $ should satisfy the condition $ \lambda_{m(d)} \le \lambda_{m(d)}^{max}=\Gamma_{m(d)}/2= \gamma_{m(d)}/2 (1+ \mathcal{C}_{m(d)})$. The interesting case is the regime with $ \mathcal{C}_{m(d)} > 1 $ where the amplification of generated Casimir photons and phonons can occur if 
\begin{eqnarray} \label{phononic coherent regime}
&& \frac{ \gamma_{m(d)}}{2} < \lambda_{m(d)} < \frac{ \gamma_{m(d)}}{2} [1+ \mathcal{C}_{m(d)}] . 
\end{eqnarray}

Analogous to the phononic damping rates, the optomechanically-induced cavity damping rate is defined as $ \kappa_{op}=-2 {\rm{Im}} \Sigma_a(\omega) $. In the on-resonance case ($ \omega=0 $), $ \kappa_{op}$ is given by 
\begin{eqnarray} \label{kapa_opt}
&& \kappa_{opt}= g^2 \frac{\gamma_m}{\frac{\gamma_m^2}{4}-\vert \lambda_m \vert^2} + G^2 \frac{\gamma_d}{\frac{\gamma_d^2}{4}-\vert \lambda_d \vert^2} \nonumber \\
&& \qquad =\kappa \left[ \frac{\mathcal{C}_0}{1-\xi_m^2} + \frac{\mathcal{C}_1}{1-\xi_d^2} \right] := \kappa \mathcal{C}_a ~ .
\end{eqnarray}
Note that if the condition (\ref{phononic coherent regime}) is satisfied then $ \kappa_{opt} <0 $ and also $ \tilde \lambda_a[\omega]  $ increases. Consequently, if the optomechanically-induced cavity damping rate becomes negative, then the effective cavity damping rate defined by $ \kappa_{\rm eff}= \kappa + \kappa_{opt}= \kappa (1+\mathcal{C}_a) $ decreases which leads to the heating of the cavity field.





\section{GENERATION OF CASIMIR PHOTONS AND PHONONS}\label{sec5}
\begin{figure}
	\includegraphics[width=9cm]{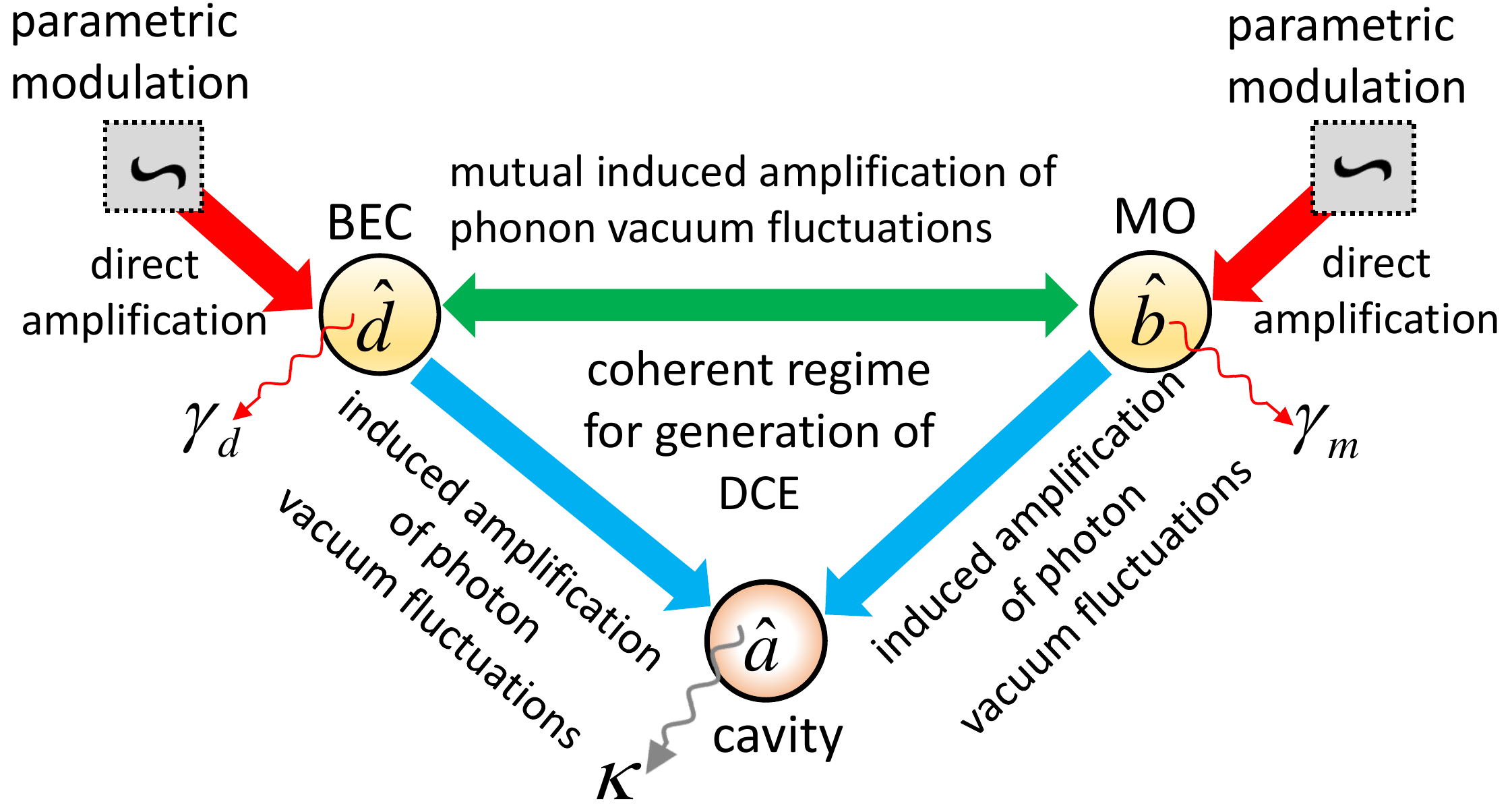}
	\caption{(Color online) A schematic diagram describing different channels for the generation and amplification of the DCE of photons and mehchanical and Bogoliubov phonons. The time modulations of the BEC and the MO establish coherent channels for induced amplification of the photon vacuum fluctuations resulting in the DCE of photons. Moreover, the time modulation of the BEC (MO) not only causes the amplification of the vacuum fluctuations of the Bogoliubov phonons (MO's phonons), but also provides a coherent channel to induce a frequency-dependent parametric amplification of the vacuum fluctuations of the MO's phonons (Bogoliubov phonons).
		 }
	\label{fig2}
\end{figure}
In this section we are going to explore the possibility of generating Casimir photons and phonons in the system under consideration through the parametric amplification of the quantum vacuum in the so-called \textit{coherent} regime, within the limit of weak optomechanical coupling ($ g,G \ll \kappa $) and the red-detuned regime of cavity optomechanics.
Moreover, it should be pointed out that although there is no direct parametric amplification term for the intracavity mode in the Hamiltonian of the system [Eq.~(\ref{H-Linear})] but due to the time modulation of the atomic collisions and the mechanical spring coefficient, a frequency-dependent amplification [Eq.~(\ref{lambda_a_omega})] can be induced to the vacuum fluctuations of the cavity mode which leads to the \textit{coherent} and the \textit{dissipative} terms corresponding to the real and imaginary parts of Eq.~(\ref{lambda_a_omega}), respectively.

In order to clarify this matter let us remind how in the previous section we showed that our system resembles a DPA with an effective frequency-dependent parametric drive (in spite of an ordinary DPA whose parametric drive is frequency-independent). It should be noted that in an ordinary DPA when its parametric drive is real the interaction is called a coherent interaction.
Generally in our system, a frequency-dependent effective parametric drive like $ \tilde \lambda_a[\omega] $ satisfies the relation $(\tilde \lambda_a[\omega])^\ast=\tilde \lambda_a^\ast[-\omega] $. 
 However, if $ \tilde \lambda_a[\omega] $ satisfies the relation $(\tilde \lambda_a[\omega])^\ast=\tilde \lambda_a[\omega] $ the system behaves like an ordinary DPA in a coherent regime. In this situation, $ \tilde \lambda_a[\omega] $ can be considered as an effective coherent interaction strength whose real and imaginary parts might lead to the \textit{`coherent'} and \textit{`dissipative'} behaviors. In this way, two different regimes can be distinguished depending on which part is larger than the other. In the so-called coherent regime where
\begin{eqnarray}  \label{coherent regime}
&&  \lvert \frac{{\rm Im} \tilde \lambda_a(\omega)}{{\rm Re} \tilde \lambda_a(\omega)} \vert \ll 1 
\end{eqnarray} 
is satistied, the \textit{coherent} term is dominant and consequently the photon generation via the DCE can be achieved. Otherwise, the \textit{dissipative} term is dominant and the photon generation via the DCE can not be amplified. The relation (\ref{coherent regime}) can be satisfied in the largely different coupling rates regime with nonzero modulation such that $ \xi_{m(d)} > 1$.


Regarding to the two phononic subsystems, i.e., the Bogoliubov and the MO modes, Eqs.~(\ref{lambda_b_omega}) and (\ref{lambda_d_omega}) show that in addition to the direct amplification of the phonon vacuum fluctuations of each subsystem due to its time modulation (the first term), there is an induced amplification (the second term) arising from the time modulation of the other subsystem indicating that there exist two channels for generation of the Casimir phonons of the BEC and the mechanical modes [see Fig.~(\ref{fig2})]. In the following, we discuss the DCE in the system under consideration for three different situations of time modulation.

\subsection{Mechanical modulation in the absence of BEC}

\begin{figure} 
	\includegraphics[width=4.29cm]{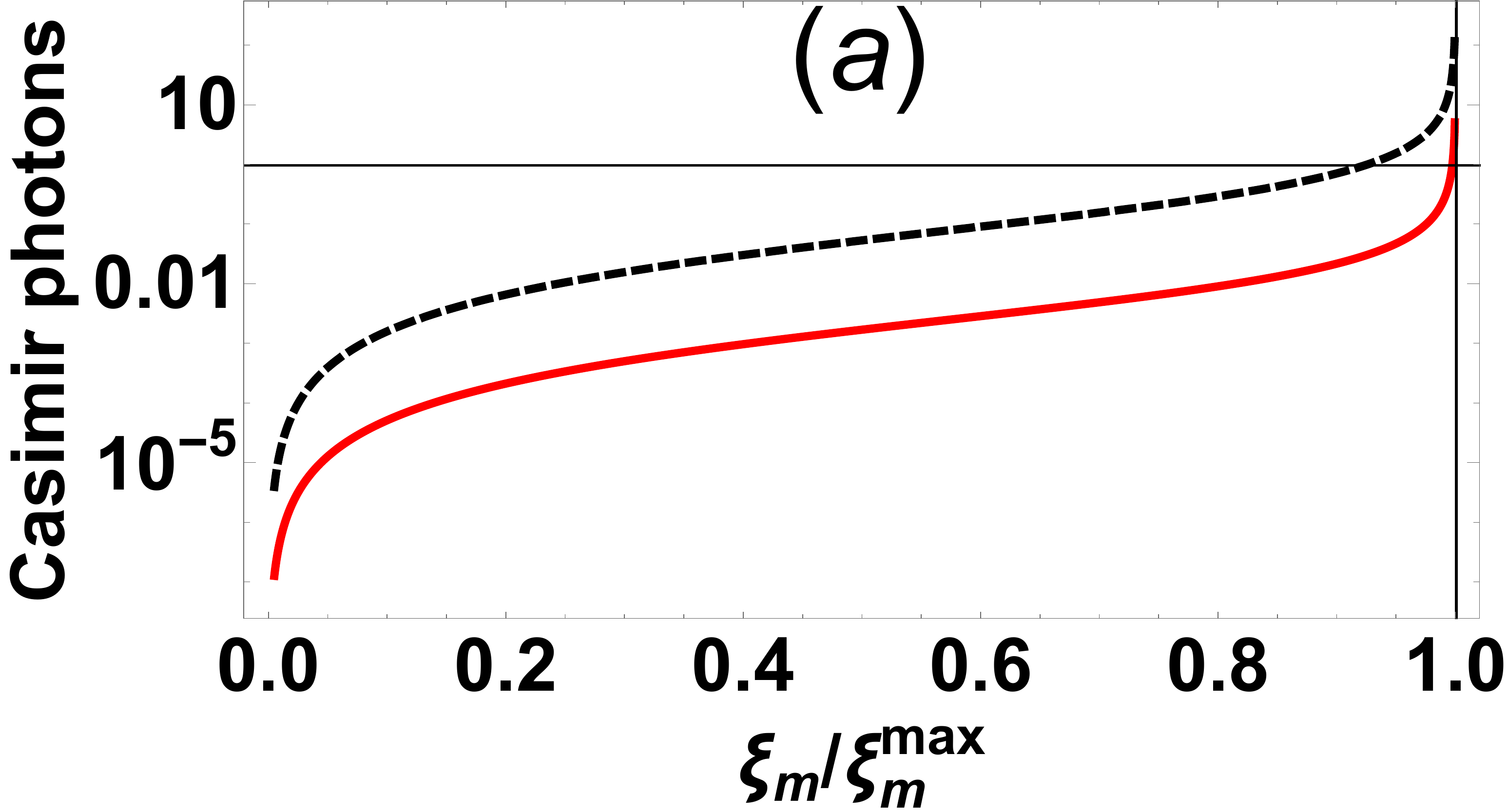}
	\includegraphics[width=4.28cm]{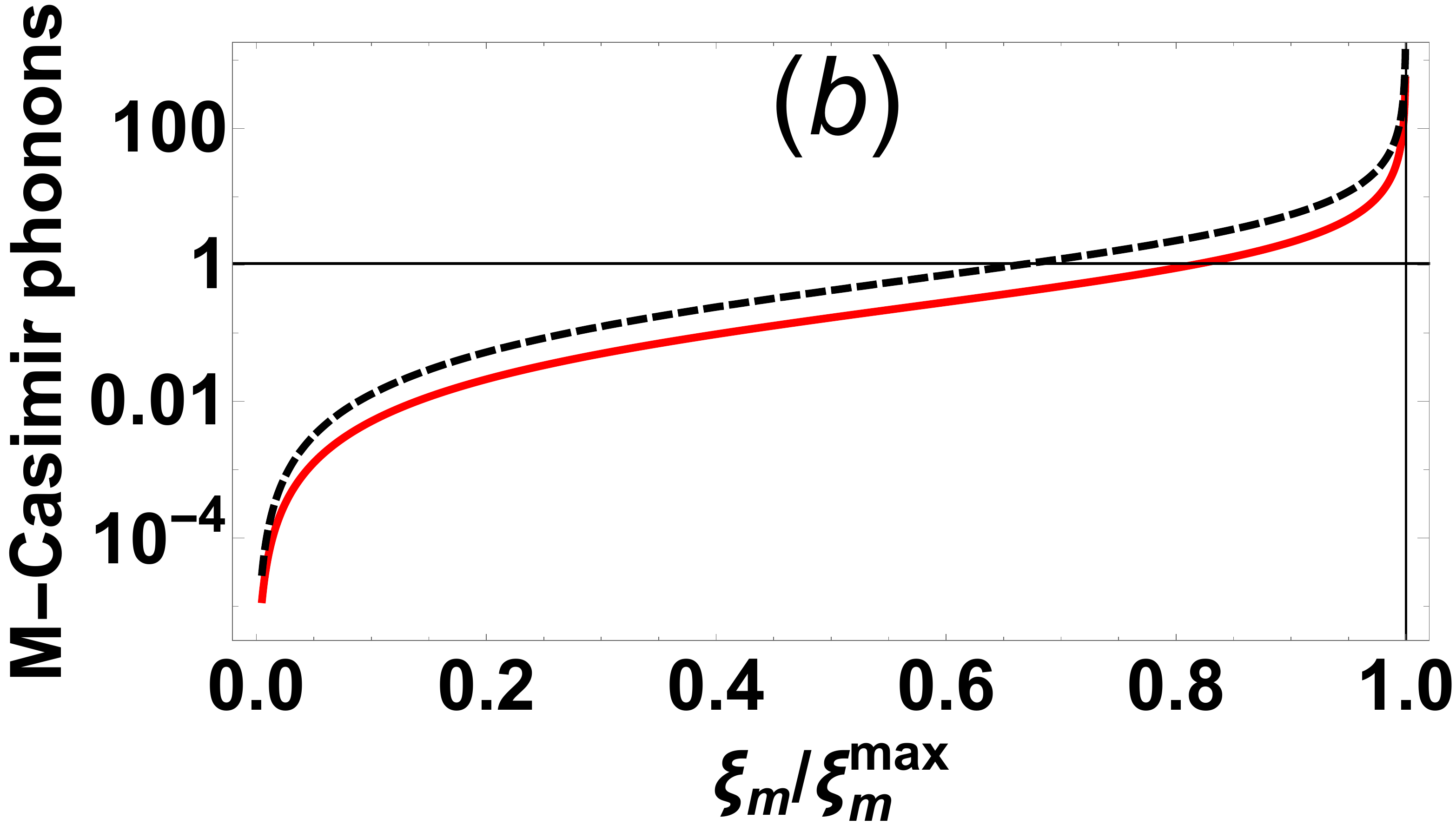}
	\caption{(Color online) (a) steady-state mean number of generated Casimir photons and (b)   mechanical-type Casimir phonons vs $ \xi_m / \xi_m^{\rm max} $ in the absence of the BEC. The red solid and black dashed lines correspond, respectively, to $ g/\kappa=0.05 $, $ \mathcal{C}_0=100 $ and $ g/\kappa=0.25 $, $ \mathcal{C}_0=2500 $. Here, we set $ \kappa/\gamma_m=10^{4} $. }
	\label{fig3}
\end{figure}

In Fig.~(\ref{fig3}), the steady-state mean number of generated Casimir photons and mechanical-type Casimir phonons in the absence of the BEC ($ G_0=\lambda_d=\gamma_d= 0$) have been plotted versus $ \xi_m / \xi_m^{\rm max} $ for a large value of cooperativity in the weak coupling regime. As the effective parameter $ \xi_m $ increases (due to the increase of modulation amplitude) to its maximum value $ 1+\mathcal{C}_m $ (in this case $ \mathcal{C}_m=\mathcal{C}_0 \gg 1 $) the mean numbers of the generated Casimir photons and phonons asymptotically increase. However, in order to generate considerable number of Casimir photons,  much more values of $ \xi_m $ is needed in comparison with that needed for mechanical-type Casimir phonons generation. 
On the other hand, by increasing the optomechanical coupling strength between the MO and the cavity field, which is equivalent to increasing the cooperativity, more number of Casimir photons and phonons can be generated at smaller values of $ \xi_m $. 
Moreover, the ratio of the damping rates, $ \kappa/\gamma_m $, can only affect the time period over which the system reaches to the steady state without any significant effect on the mean number of the generated Casimir particles at the steady state. It should be mentioned that for the parameter values given in Fig.~(\ref{fig3}), the mechanical and optical vacuum fluctuations start to be amplified for $ \xi_m / \xi_m^{\rm max} \ge 0.5 $ and 0.9, respectively.

\subsection{Modulating atomic collisions without MO modulation}
\begin{figure} 
	\includegraphics[width=4.29cm]{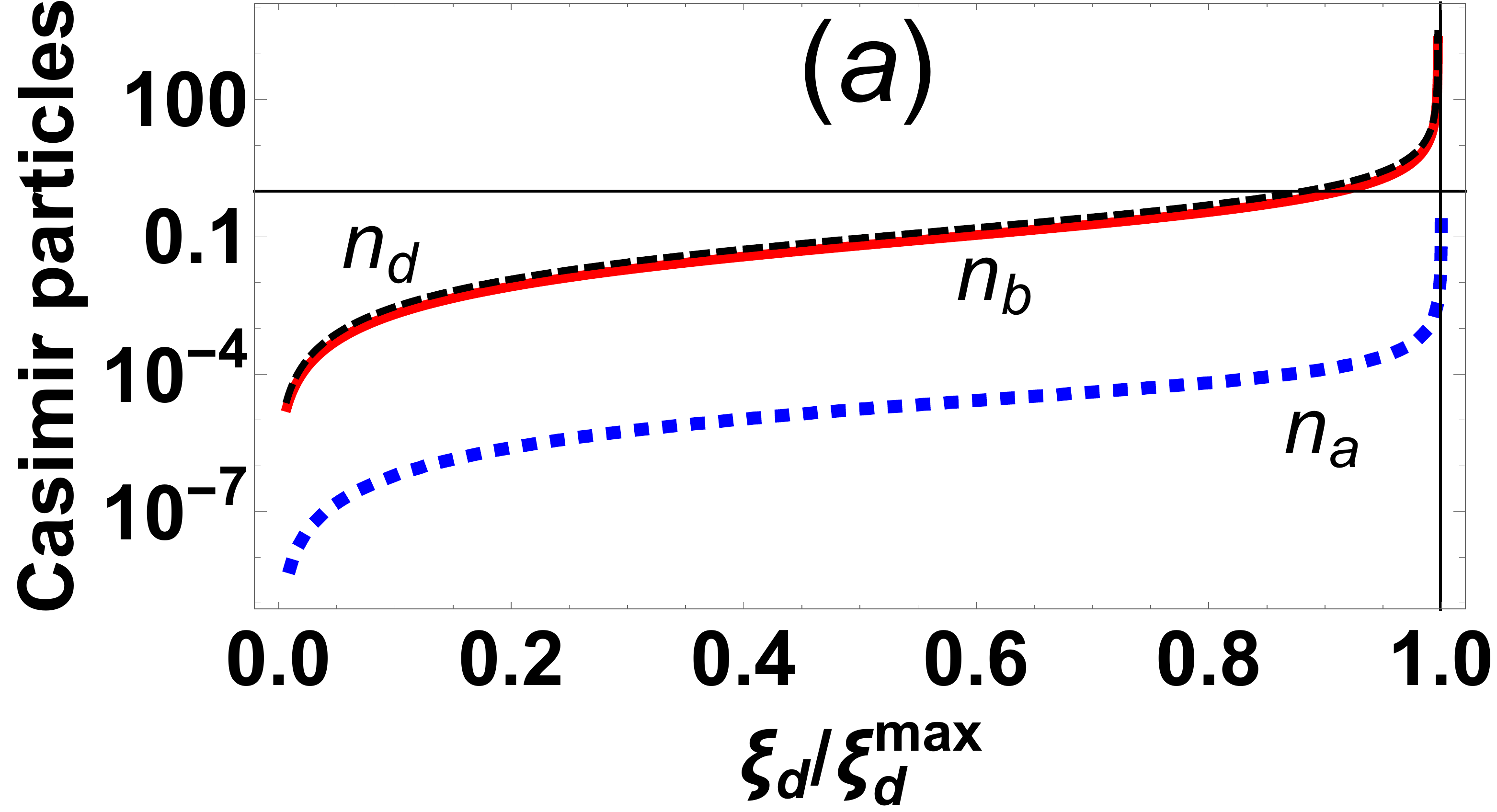}
	\includegraphics[width=4.28cm]{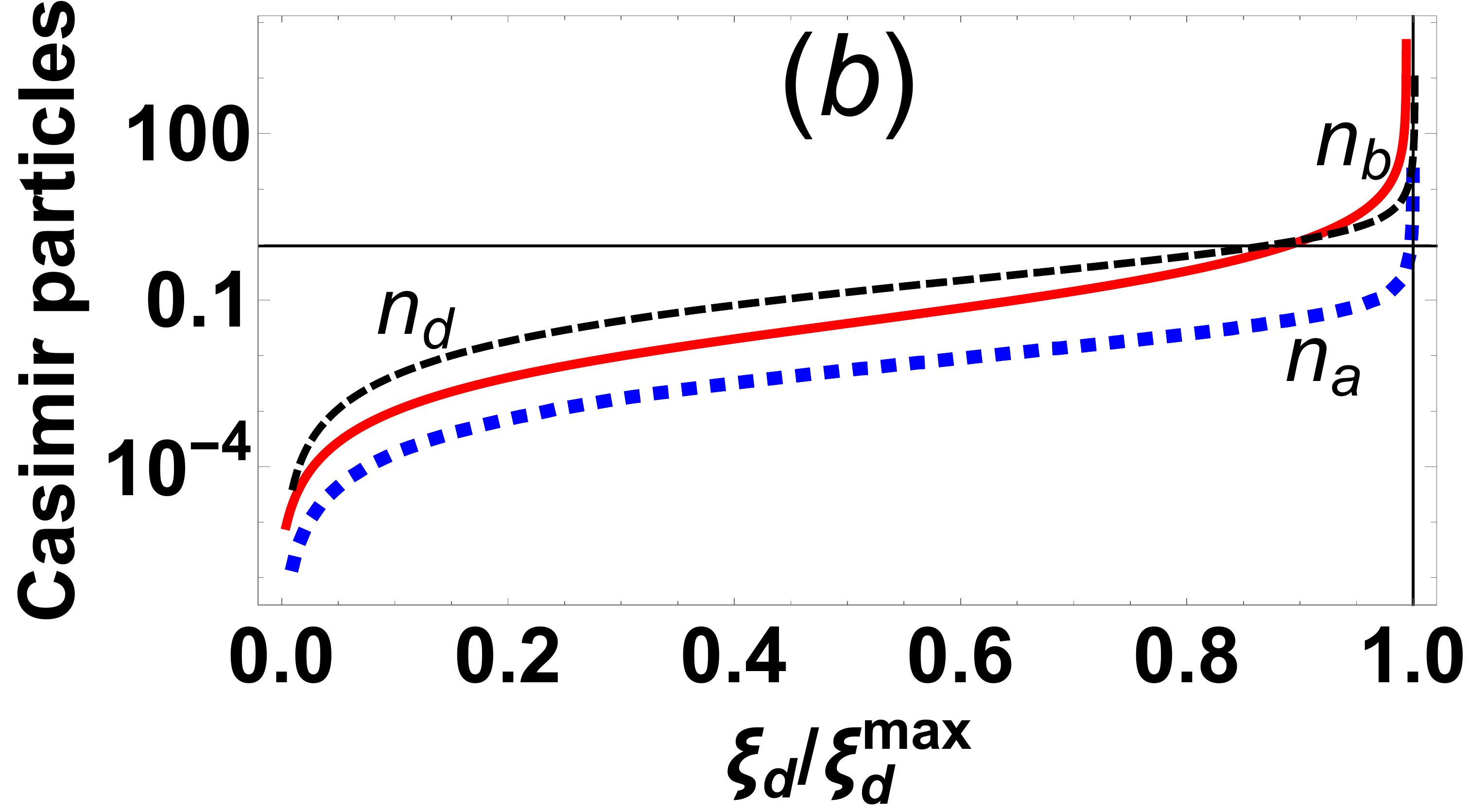}
	\caption{(Color online) Steady-state mean number of the generated Casimir photons and phonons vs $ \xi_d/\xi_d^{max} $ when mechanical modulation is zero ($ \xi_m=0 $) for the cases of (a) equal optomechanical coupling strengths $ g=G=0.05 \kappa $ and (b) largely different optomechanical coupling strength $ G=20g=0.1 \kappa $. The dashed black, solid red, and dotted blue lines correspond, respectively, to the Bogoliubov-type Casimir phonons, mechanical-type Casimir phonons, and Casimir photons. Here, we have set $ \gamma_m=\gamma_d=10^{-3}\kappa $.}
	\label{fig4}
\end{figure}

In Fig.~(\ref{fig4}) the mean number of the generated Casimir photons and mechanical/Bogoliubov-type phonons have been plotted  versus $ \xi_d / \xi_d^{\rm max} $ in the weak coupling regime ($ g,G\ll \kappa $) and in the absence of mechanical modulation ($ \xi_m=0 $). As shown in Fig.~(\ref{fig4}a), in the regime of equal optomechanical coupling strengths of the MO and the BEC with the cavity mode, $ g=G $, the steady-state mean number of the generated Casimir photons is negligibly small over the entire range of $ \xi_d / \xi_d^{\rm max} $ which means that equal couplings is corresponding to the dissipative regime for the intracavity field. However, the mean number of the generated mechanical/Bogoliubov-type Casimir phonons is considerably increased for $ \xi_d / \xi_d^{\rm max}> 0.8 $ and grows up asymptotically as $ \xi_d $ reaches to its maximum value.

On the other hand, as shown in Fig.~(\ref{fig4}b), in the regime of largely different optomechanical coupling strengths ($ \vert g-G \vert\gg g  $) which is equivalent to the regime of largely different cooperativities, Casimir photons as well as Casimir phonons can be generated considerably by increasing the atomic collisions modulation parameter $ \xi_d $ which means that this regime corresponds to the coherent regime for the intracavity light field.
Therefore, by controlling the modulation rate of atomic collisions one can control the number of Casimir photons and phonons. Furthermore, our numerical calculations show that by increasing the difference between the optomechanical coupling rates $ g $ and $ G $ both the photonic and phononic DCEs are strengthened.  Moreover, increasing the coupling rates $ g $ and  $ G $ leads to the increase of Bogoliubov- and mechanical-type Casimir phonons, respectively.  

\begin{figure*} 
	\includegraphics[width=5.85cm]{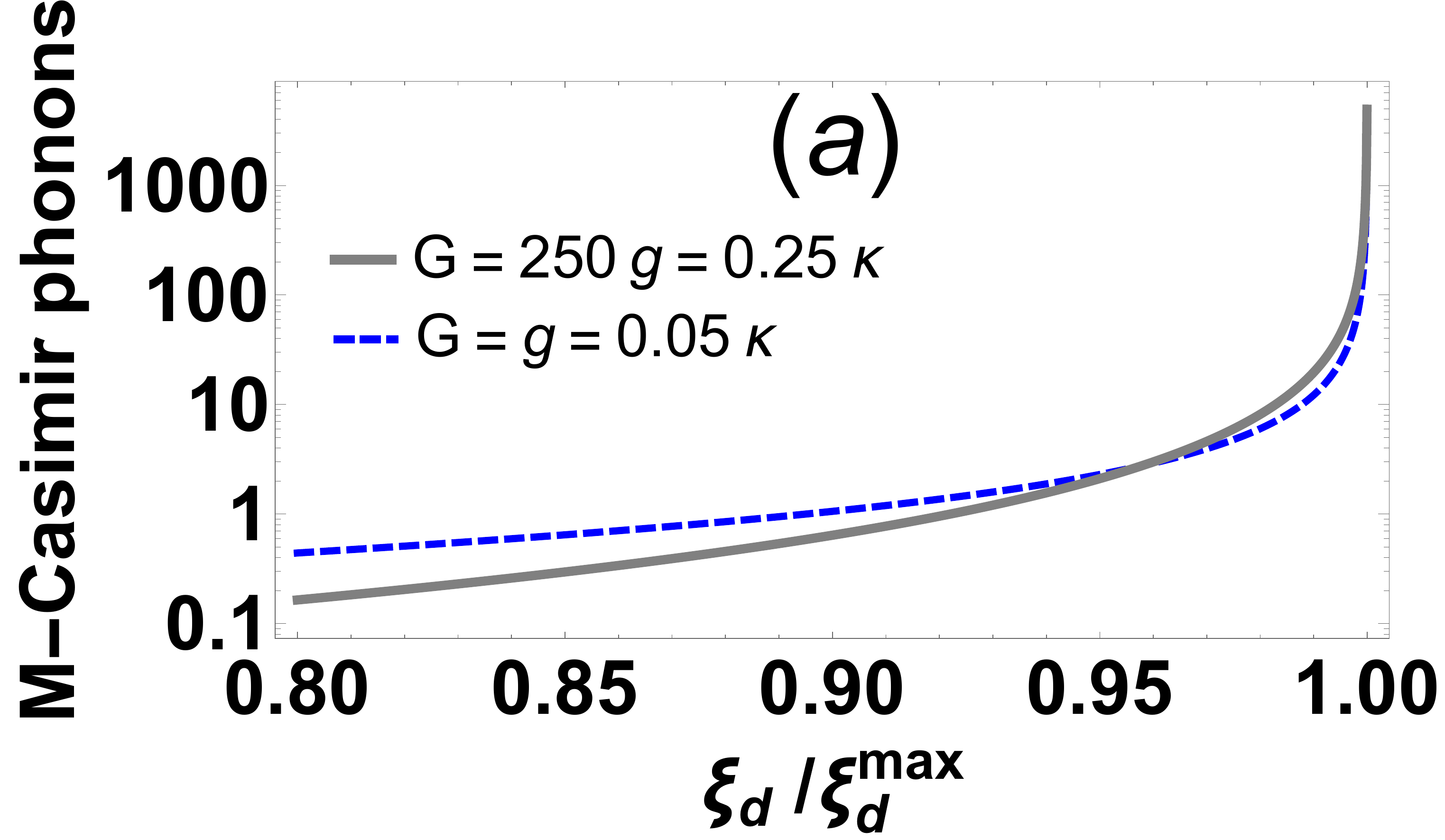}
	\includegraphics[width=5.85cm]{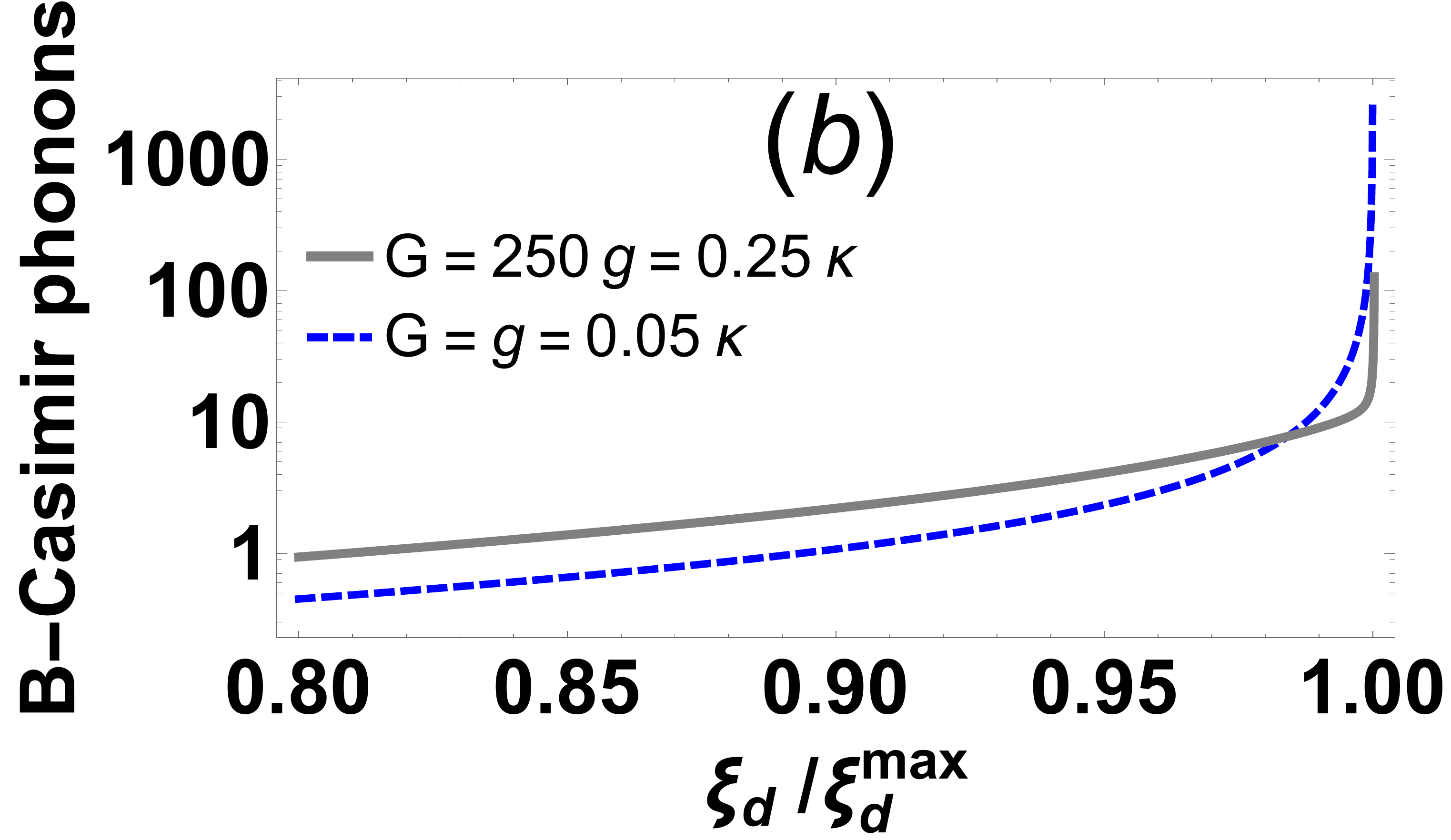}
	\includegraphics[width=5.85cm]{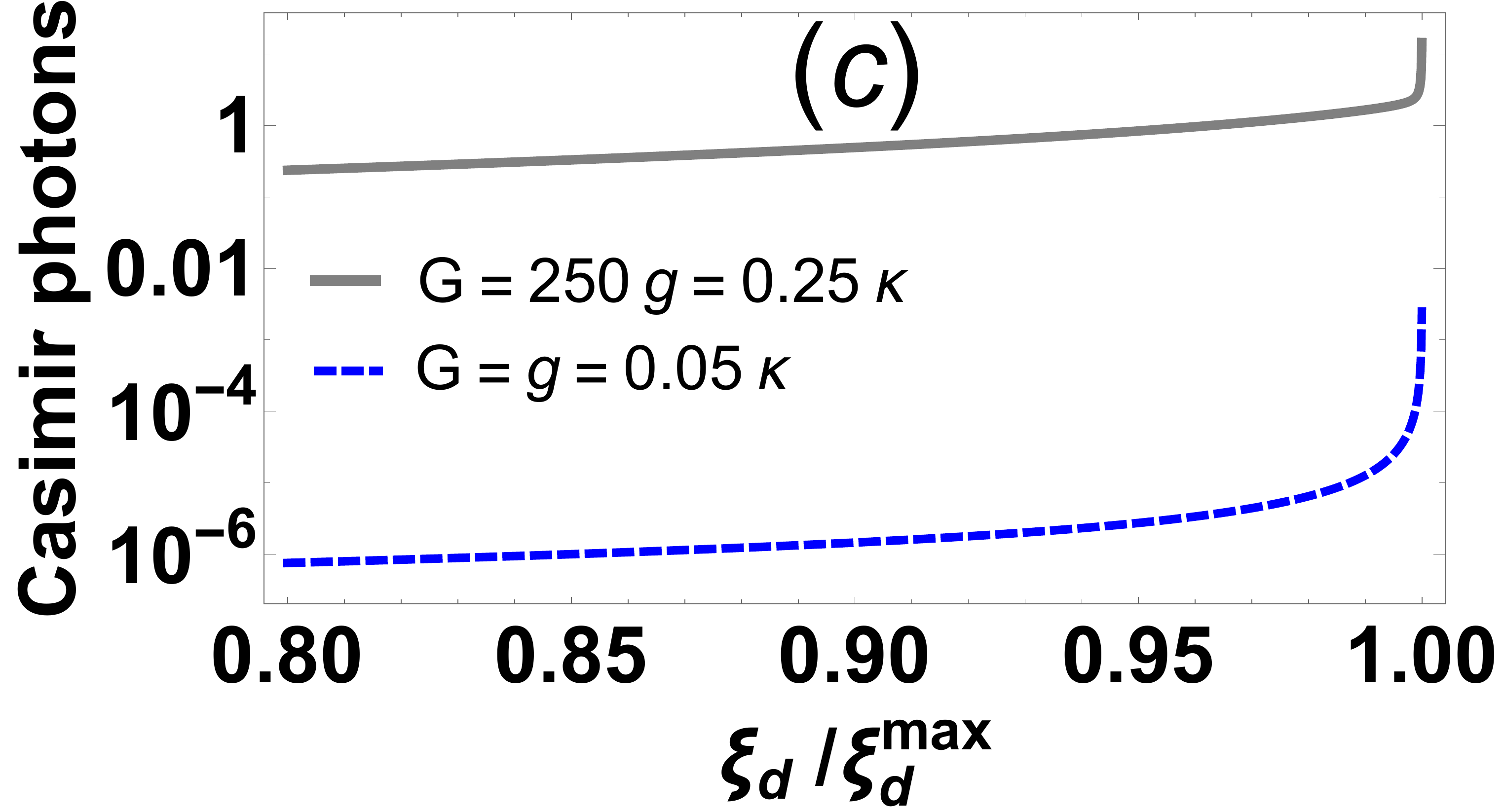}
	\caption{(Color online) Comparison between the steady-state mean numbers of generated Casimir particles in the two regimes of equal optomechanical couplings $ g=G=0.05\kappa $ (blue dashed line) and largely different one $ G=250g=0.25\kappa $ (gray solid line) in the absence of the time modulation of the MO ($ \xi_m=0$): (a) mean number of generated mechanical-type Casimir phonons, (b) mean number of Bogoliubov-type Casimir phonons, and (c) mean number of intracavity Casimir photons vs $ \xi_d/\xi_d^{max} $. Here, we have set $ \gamma_m/\kappa=\gamma_d/\kappa=10^{-4}$.
	} 
	\label{fig5} 
\end{figure*}

Let us illustrate why in our system the strong amplification of vacuum fluctuations is occurred in the regime of largely different optomechanical coupling strengths (specially for photons). Recently, it has been shown \cite{JLiNJP,JLiPRA} that in the absence of parametric modulation, the dynamics of an optomechanical system with two mechanical modes is governed by the effective linearized Hamiltonian  $ \hat H_{\rm eff}/\hbar = \Delta_0 \delta \hat a^\dag \delta \hat a + \mathcal{G}(\hat B \delta \hat a^\dag + \hat B^\dag \delta \hat a) $ where $ \hat B= (g \delta \hat b - G \delta \hat d)/\mathcal{G} $ denotes a collective mode and $ \mathcal{G}=\sqrt{G^2-g^2} $ describes the rate of excitation exchange between the collective mode $ \hat B $ and the cavity field which leads to the cooling of the collective mode. The collective mode $ \hat B $ can approach its vacuum state corresponding to a two-mode squeezed state of the two mechanical modes. 
In our system the Bogoliubov mode of BEC plays formally the role of a second mechanical oscillator mode and the two oscillators (MO and BEC) are driven parametrically. However, as has been shown in Ref. \cite{dalafi5} , by a suitable unitary transformation our system Hamiltonian can be transformed into an ordinary optomechanical one without any parametric modulation (similar to the above-mentioned Hamiltonian). Therefore, we can also construct a similar collective mode $\hat  B $ out of the atomic and the mechanical modes.
Note that, in general, one can consider the collective mode as a Bogoliubov-transformed mode  $ \hat B=  \delta \hat b \cosh r -  \delta \hat d \sinh r  $ with the squeezing parameter  $ r= \tanh^{-1} (g/G) $. Therefore, the ratio $ g/G $ determines how much the vacuum of the collective mode is squeezed or amplified. In particular, the vacuum of the collective mode in the limit of equal optomechanical coupling strengths ($  g/G \to 1 $) corresponds to a maximally squeezed or amplified state. However, in this limit the collective mode is decoupled from the optical mode (because $ \mathcal{G} \to 0 $), and thus the Casimir photons can no longer be amplified. That is why in our system the Casimir photons are strongly generated in the regime of largely different optomechanical coupling strengths (or equivalently, largely different cooperativities) where the collective mode $ \hat B $ interacts with the optical mode effectively.

In order to see more clearly the effects of the two regimes of equal and largely different optomechanical couplings on the steady-state mean numbers of generated Casimir photons and phonons, we have plotted the number of phonons [Fig.~(\ref{fig5}a) for mechanical-type and Fig.~(\ref{fig5}b) for Bogoliubov-type phonons] and the number of photons [Fig.~\ref{fig5}(c)] versus $ \xi_d / \xi_d^{\rm max} $ for the two mentioned regimes. As is seen in Fig. (\ref{fig5}c),  the number of generated Casimir photons in the regime of largely different couplings is always larger than that in the equal coupling strength regime which is acceptable based on the collective mode interpretation mentioned in the previous paragraph. Also, as the difference between the two couplings increases, the coherent channels or coherent regime for the intracavity light field is dominant. Consequently, we can conclude that the increases of the difference between coupling rates and of the amplitude of modulation of the atomic collision frequency are the only two important factors to amplify the generation of the Casimir photons.

On the other hand, the behavior of the generated phonons are somehow different from that of photons. As is seen from Figs.~(\ref{fig5}b) and (\ref{fig5}c), for the mechanical/Bogoliubov mode there exists a critical value of $\xi_{d}$ so that the behavior of the generated Bogoliubov or mechanical-type Casimir phonons are different below and above it. Figure (\ref{fig5}b) shows that below the critical value the number of Bogoliubov phonons in the regime of largely different optomechanical couplings is larger than that in the regime of equal couplings while the situation is reversed above the critical value. It can be interpreted using collective phononic mode in terms of $ \mathcal{G} $ which means that there is an optimum value of $ G/g $ or $ \xi_{d,m} $ to achieve the maximum number of generated phonons. However, as is seen from Fig.~(\ref{fig5}a), the behavior of the generated mechanical-type Casimir phonons is reverse in comparison with that of the Bogoliubov-type Casimir phonons. It is because of the fact that here the mechanical modulation is off ($ \xi_{m}=0 $) while the atomic modulation which drives the Bogoliubov mode is on ($ \xi_{d} \neq 0 $).

Note that in the absence of the MO modulation, the mean number of generated Casimir photons and phonons can be controlled by the effective modulation of the atomic collisions frequency ($ \xi_d $) and also the difference between the coupling rates of the cavity light field to the BEC and the MO. It should be mentioned that the same results can be obtained for the case without BEC modulation when only the MO modulation is present since the Bogoliubov mode in the BEC plays formally the role of a second MO mode.

\subsection{Simultaneous modulation of both the MO and atomic collisions}
\begin{figure*} 
	\includegraphics[width=5.87cm]{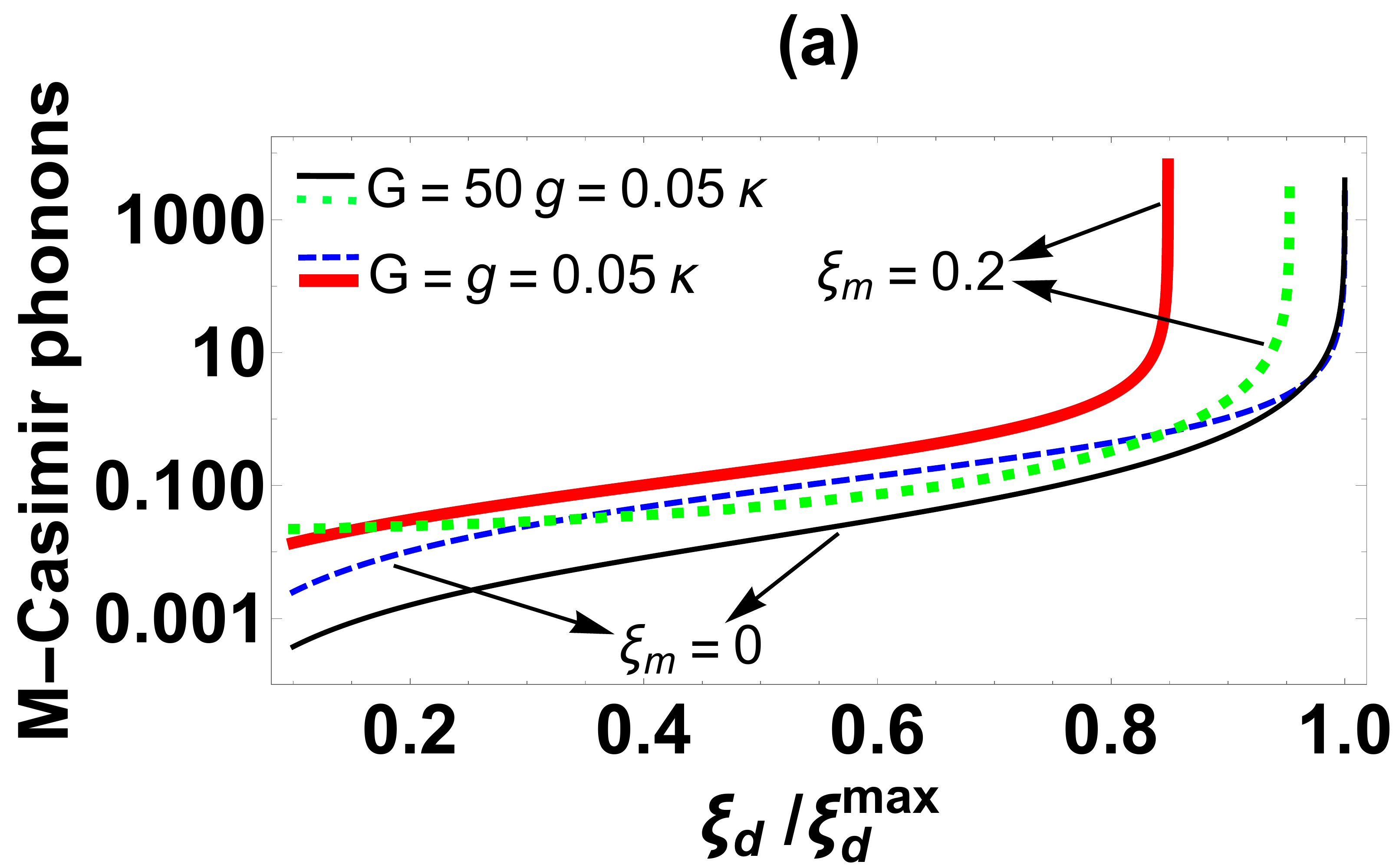}
	\includegraphics[width=5.95cm]{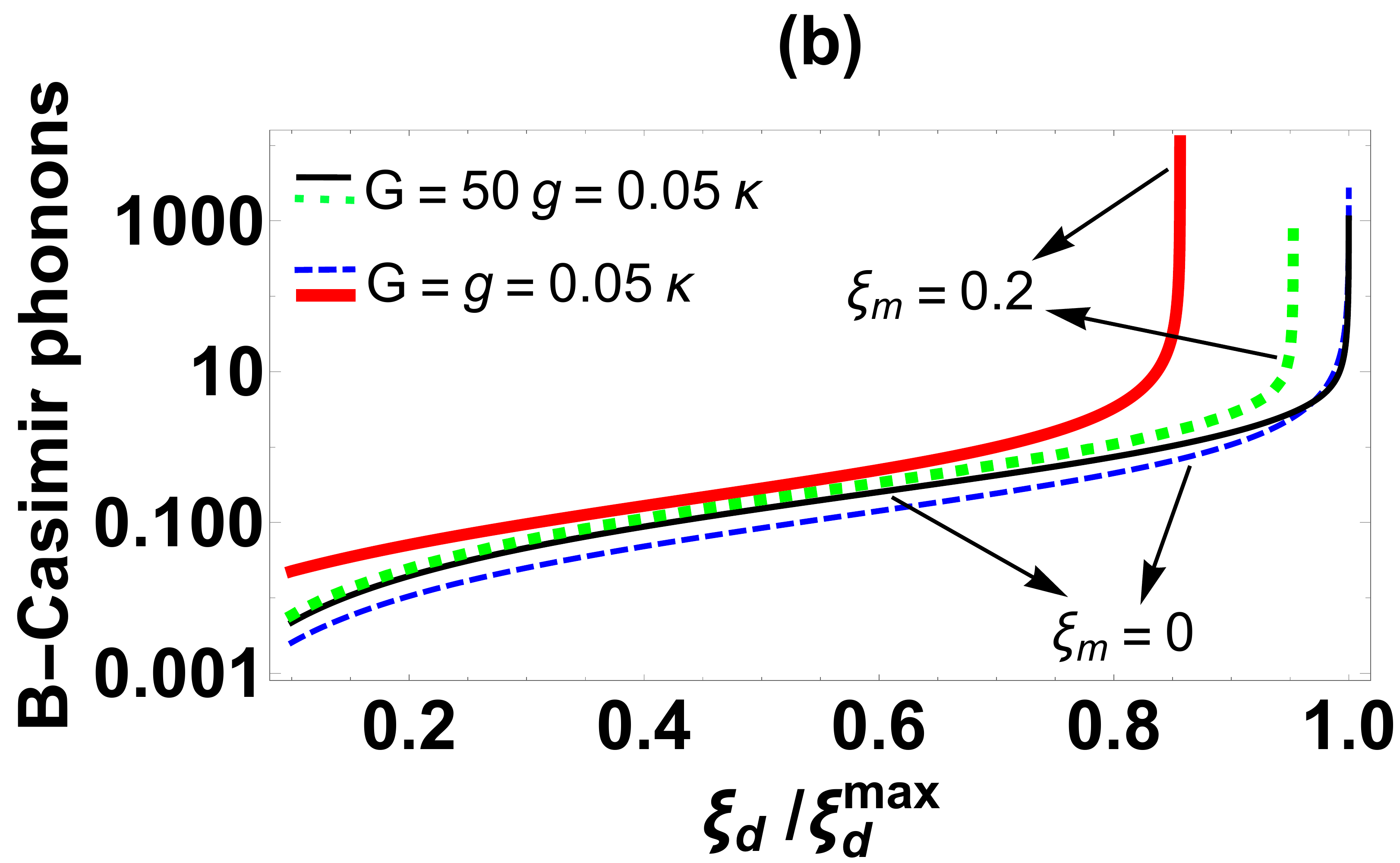}
	\includegraphics[width=5.95cm]{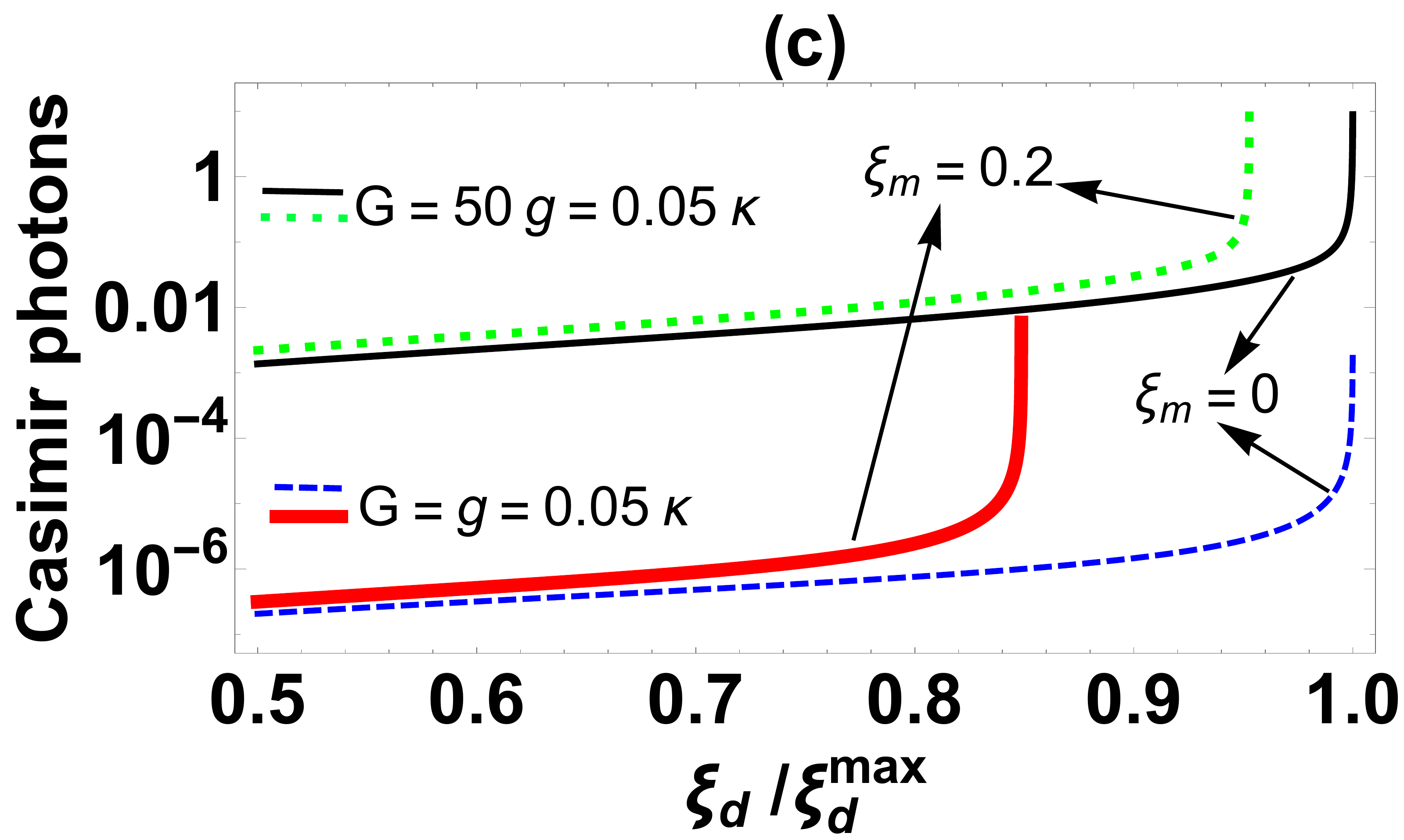}
	\caption{(Color online) steady-state mean number of generated Casimir radiations vs $ \xi_d/\xi_d^{max} $ in equal couplings regime corresponding to $ g=G=0.05 \kappa $ (red thick and blue dashed lines) and largely different couplings regime corresponding to $ G=50g=0.05 \kappa $ (black thin and green dotted lines) for two values of the mechanical parametric modulation $ \xi_m=0 $ (black thin and blue dashed lines) and $ \xi_m=0.2 $ (red thick and green dotted lines). (a) mechanical-type Casimir phonons, (b) Bogoliubov-type Casimir phonons, and (c) Casimir photons. Here, we set $ \gamma_m=\gamma_d=10^{-4}\kappa $. } 
	\label{fig6}
\end{figure*}

Figure (\ref{fig6}) shows the steady-state mean number of generated Casimir phonons and photons versus $ \xi_d / \xi_d^{max} $ in the weak coupling regime and in the presence of both the BEC and MO modulations. As is expected, in the equal optomechanical couplings regime, the Casimir photons cannot be generated [Fig.~(\ref{fig6}c)] while a considerable number of the mechanical/Bogoliubov-type Casimir phonons are generated and increased by increasing $ \xi_d $ [Figs. (\ref{fig6}a) and (\ref{fig6}b)]. As is evident from Fig. (\ref{fig6}) by turning on the MO modulation the mean number of generated photons and phonons increase for each value of $ \xi_d $. A similar interpretation like that was explained in the previous subsection is valid here in the presence of both modulations of the atomic collisions frequency and the mechanical spring coefficient.

Since $ \lambda_m $ depends on the modulation amplitude of the MO  while $ \lambda_d $ depends both on the modulation amplitude of the BEC and on the atomic collisions frequency $ \omega_{sw} $, the mean number of generated Casimir photons and phonons can be controlled externally not only through the modulation parameters but also through the atomic collisions frequency which itself is experimentally controllable by the transverse trapping frequency $ \omega_{\perp} $. Also, through the input power laser and intrinsic coupling rates, one can control the regime corresponding to the generation of the Casimir particles, e.g., coherent regime for the photonic and phononic subsystems. Therefore, in contrast to other DCE proposals for photons or phonons \cite{Daviesbook.Dodonov2.Dalvit.Dodonov3.Crocce.Dodonov4.Dodonov5}, our system is more controllable. Moreover, similar to the previous subsection, the same results can be obtained by exchanging the roles of the BEC and MO modulations.

On the other hand, it has recently been shown \cite{antiDCE-work} that the coherent generation and annihilation of photons, which are respectively corresponding to the DCE and \textit{anti}-DCE in the quantum Rabi model with time-modulated atomic frequency in the dispersive, moderate coupling and weak modulation regimes \cite{Motazedifard DCE,antiDCE}, leads to the positive and negative quantum work extraction in the context of quantum \textit{thermodynamics}.  Also, it has been shown that a realistic out-of-equilibrium finite-time protocol harnessing anti-DCE allows for work extraction. The effect of the photon creation in our system can be attributed to the exchange of energy between the external modulation of the BEC/MO and the quantum vacuum fluctuations of the photonic and phononic subsystems through the coherent channels. If the energy is transferred to the quantum vacuum fluctuations, Casimir pair particles are created. Therefore, based on Ref. \cite{antiDCE-work}, one can conclude that the positive quantum work in our system can be extracted and externally controlled by the modulation parameters of the atomic collisions frequency and the mechanical spring coefficient which open a new way to the quantum thermodynamics of the DCE physics.

\section{Summary and outlooks}\label{sec6}
In summary, we have theoretically proposed and investigated a feasible experimental scheme to realize the DCE of phonons and photons in a hybrid optomechanical cavity with a moving end mirror containing an interacting cigar-shaped BEC. It is shown that by coherent modulation of the \textit{s}-wave scattering frequency of the BEC and of the mechanical spring coefficient of the moving mirror, the mechanical and atomic vacuum fluctuations are parametrically amplified which consequently lead to the generation of the mechanical- and Bogoliubov-type Casimir phonons. Also, in the regime of largely different optomechanical couplings, the quantum vacuum fluctuations of the intracavity field are indirectly amplified via the coherent channels by the MO and BEC which lead to the generation of cansiderable number of Casimir photons in comparison to the conventional DCE proposals.

One of the important advantages of the present scheme in comparison to the other proposals is its controllability. As was shown, the mean number of generated Casimir photons and phonons can be controlled externally not only through the modulation parameters but also through the atomic collisions frequency of the BEC which itself is experimentally controllable by the transverse trapping frequency.

As some interesting outlooks of the present work, one can investigate the possibility of the perfect quadrature squeezing or quadrature amplification of the MO/BEC or cavity output. Moreover, it seems that in the system under consideration due to the generation of the Casimir photons and phonons one can require the condition for simultaneously backaction noise suppression and signal amplification in order to perform single quadrature force sensing in both on- and off-resonance frequency by controlling the atomic collisions frequency. Furthermore, one can look for the condition under which the cavity spectral photon function becomes negative while the system is stable, which leads to identifying an externally controllable negative effective temperature for photons that can be responsible for the modified optomechanically induced transparency (OMIT). Finally, it seems that our proposed scheme has the potential to generate robust entanglement between the MO mode and the Bogoliubov mode of the BEC which plays formally the role of a second MO mode. We hope to report on these issues in the near future.

\textit{\textbf{Acknowledgements}}
A. M. wishes to thank the Office of Graduate Studies of the University of Isfahan and also the Iran's National Elites Foundation (INEF).
A. D. wishes to thank the Laser and Plasma Research Institute of Shahid Beheshti University for their support.	


\begin{thebibliography}{60}



\bibitem{Schwinger} J. Schwinger, 
 \href{https://doi.org/10.1103/PhysRev.82.664}{Phys. Rev. \textbf{82}, 664 (1951)}.

\bibitem{Hawking} S. W. Hawking, 
 \href{}{Nature \textbf{248}, 30(1974)}; 
S. W. Hawking, 
 \href{https://doi:10.1007/BF02345020}{Commun. Math. Phys. \textbf{43}, 199 (1975)}.

\bibitem{Unruh} W. G. Unruh, 
 \href{https://doi.org/10.1103/PhysRevD.14.870}{ Phys. Rev. D \textbf{14}, 870 (1976)}.

\bibitem{Yablonovitch} E. Yablonovitch, 
 \href{https://doi.org/10.1103/PhysRevLett.62.1742}{Phys.Rev. Lett. \textbf{62}, 1742 (1989)}.

\bibitem{Schwinger2} J. Schwinger, 
 \href{https://doi.org/10.1073/pnas.89.9.4091}{Proc. Natl Acad. Sci. USA \textbf{89}, 4091 (1992)}.

\bibitem{Wilson} J. R. Johansson, G. Johansson. C. M. Wilson, and F. Nori, 
 \href{https://doi.org/10.1103/PhysRevA.82.052509}{ Phys. Rev. A \textbf{82}, 052509 (2010)}.

\bibitem{Moore} G. T. Moore, 
 \href{https://doi.org/10.1063/1.1665432}{J. Math. Phys. \textbf{11}, 2679 (1970)}.

\bibitem{Davies} P. C. W. Davies and S. A. Fulling, 
 \href{https://doi.org/10.1098/rspa.1977.0130}{Proc. Roy. Soc. London. A \textbf{356}, 237 (1977)}.

\bibitem{Dodonov1} V. V. Dodonov, 
 \href{https://doi.org/10.1088/0031-8949/82/03/038105}{Phys. Scripta \textbf{82}, 038105 (2010)}.

\bibitem{Nation} P. D. Nation, J. R. Johansson, M. P. Blencowe, and F. Nori, 
 \href{https://doi.org/10.1103/RevModPhys.84.1}{Rev. Mod. Phys. \textbf{84}, 1 (2012)}.




\bibitem{Daviesbook.Dodonov2.Dalvit.Dodonov3.Crocce.Dodonov4.Dodonov5} N. D. Birrell and P. C. D. Davies, \textit{Quantum Fields in Curved Space} \href{https://doi.org/10.1017/CBO9780511622632}{(Cambridge University, 1982)};
V. V. Dodonov, 
 \href{https://doi.org/10.1103/PhysRevA.58.4147}{Phys. Rev. A \textbf{58}, 4147 (1998)}.
 D. A. R Dalvit and F. D. Mazzitelli, 
  \href{https://doi.org/10.1103/PhysRevA.59.3049}{Phys. Rev. A \textbf{59}, 3049 (1999)}.
 V. V. Dodonov, 
  \href{https://doi.org/10.1002/0471231479.ch7}{Adv. Chem. Phys. \textbf{119}, 309 (2001)}.
 M. Crocce, D. A. R. Dalvit, and F. D. Mazzitelli, 
  \href{https://doi.org/10.1103/PhysRevA.66.033811}{Phys. Rev. A \textbf{66}, 033811 (2002)}.
 A. V. Dodonov, E. V. Dodonov, and V. V. Dodonov, 
  \href{https://doi.org/10.1016/j.physleta.2003.08.065}{Phys. Lett. A \textbf{317}, 378 (2003)}.
 A. V. Dodonov, R. Lo Nardo, R. Migliore, A. Messina, and V. V. Dodonov, 
  \href{https://doi.org/10.1088/0953-4075/44/22/225502}{J. Phys. B: At. Mol. Opt. Phys. \textbf{44}, 225502 (2011)}.
 
\bibitem{Dodonov6} V. V. Dodonov, A. B. Klimov, and V. I. Manko, 
 \href{https://doi.org/10.1016/0375-9601(90)90333-J}{Phys. Lett. A \textbf{149}, 225 (1990)}.

\bibitem{Dodonov7} V. V. Dodonov and M. A. Andreata, 
 \href{https://doi.org/10.1088/0305-4470/32/39/301}{J. Phys. A \textbf{32}, 6711 (1999)}.

\bibitem{Johansson} J. R. Johansson, G. Johansson, C. M. Wilson, P. Delsing, and F. Nori, 
 \href{https://doi.org/10.1103/PhysRevA.87.043804}{Phys. Rev. A \textbf{87},043804 (2013)}.

\bibitem{Bhattacherjee} N. Aggarwal, A. B. Bhattacherjee, A. Banerjee, and M. Mohan, 
 \href{https://doi.org/10.1088/0953-4075/48/11/115501}{J. Phys. B: At. Mol. Opt. Phys.\textbf{48}, 115501 (2015)}.

\bibitem{Felicetti} Felicetti, M. Sanz, L. Lamata, G. Romero, G. Johansson, P. Delsing, and E. Solano, 
 \href{https://doi.org/10.1103/PhysRevLett.113.093602}{Phys. Rev. Lett. \textbf{113}, 093602 (2014)}.

\bibitem{Sabin} C. Sabin and G. Adesso, 
 \href{https://doi.org/10.1103/PhysRevA.92.042107}{Phys. Rev. A \textbf{92}, 042107 (2015)}.


\bibitem{Barton} G. Barton and C. Eberlein, Ann. Phys. 227,222 (1993).  A. Lambrecht, M. T. Jaekel, and S. Reynaud, \href{https://doi.org/10.1103/PhysRevLett.77.615}{ Phys. Rev. Lett. \textbf{77}, 615 (1996)}.


\bibitem{DodonovPDCE} V. V. Dodonov, \href{https://doi.org/10.1088/0031-8949/82/03/038105}{Phys. Scr. \textbf{82}, 038105 (2010)}. 



\bibitem{high mechanical frequency} A. D. O’Connell, M. Hofheinz, M. Ansmann, R. C. Bialczak, M. Lenander, E. Lucero, M. Neeley, D. Sank, H. Wang, M. Weides, 
 \href{https://doi:10.1038/nature08967}{Nature \textbf{464}, 697(2010)}.

\bibitem{Lombardi} C. Braggio, G. Bressi, G. Carugno, C. Del Noce, G. Galeazzi, A. Lombardi, A. Palmieri, G. Ruoso, and D. Zanello, 
 \href{https://doi.org/10.1209/epl/i2005-10048-8}{Europhys. Lett. \textbf{70}, 754(2005)}.

\bibitem{Dodonov8} V. V. Dodonov and A. V. Dodonov, 
 \href{https://doi.org/10.1088/0953-4075/39/15/S20}{J. Phys. B: At. Mol. Opt. Phys. \textbf{39}, S749 (2006)}.

\bibitem{Dodonov9} V. V. Dodonov and A. V. Dodonov, 
 \href{https://doi.org/10.1088/0305-4470/39/21/S18}{J. Phys. A: Math. Gen. \textbf{39}, 6271(2006)}.

\bibitem{Dezael} F. X. Dezael and A. Lambrecht, 
 \href{https://doi.org/10.1209/0295-5075/89/14001}{Europhys. Lett. \textbf{89}, 14001 (2010)}.

\bibitem{Faccio} D. Faccio and I. Carusotto, 
 \href{https://doi.org/10.1209/0295-5075/96/24006}{Europhys. Lett. \textbf{96}, 24006 (2011)}.

\bibitem{Motazedifard DCE} A. Motazedifard, M. H. Naderi, and R. Roknizadeh, 
 \href{https://doi.org/10.1364/JOSAB.32.001555}{J. Opt. Soc. Am. B \textbf{32}, 1555(2015)}.

\bibitem{Hoeb} F. Hoeb, F. Angaroni, J. Zoller, T. Calarc, G. Strini, S. Montangero, and G. Benenti, 
 \href{https://doi.org/10.1103/PhysRevA.96.033851}{Phys. Rev. A \textbf{96}, 033851 (2017).}.

\bibitem{Agnesi} A. Agnesi, C. Braggio, G. Bressi, G. Carugno, F. Della Valle, G. Galeazzi, G. Messineo, F. Pirzio, G. Reali, G. Ruoso, D. Scarpa, and D. Zanello, 
 \href{https://doi.org/10.1088/1742-6596/161/1/012028}{J. Phys.: Conf. Series \textbf{161}, 012028 (2009)}.

\bibitem{Kawakubo} T. Kawakubo and K. Yamamoto, 
 \href{https://doi.org/10.1103/PhysRevA.83.013819}{Phys. Rev. A \textbf{83}, 013819 (2011)}.

\bibitem{Pourkabirian} C. M. Wilson, G. Johansson, A. Pourkabirian, M. Simoen, J. R. Johansson, T. Duty, F. Nori, and P. Delsing, 
 \href{https://doi.org/10.1038/nature10561}{Nature(London) \textbf{479}, 376 (2011)}.

\bibitem{Lahteenmaki} P. , G. S. Lahteenmaki, Paraoanu, J. Hassel, and P. J. Hakonen, 
 \href{https://doi.org/10.1073/pnas.1212705110}{Proc. Natl. Acad. Sci. U.S.A. \textbf{110}, 4234 (2013)}.






\bibitem{Recati} I. Carusotto, R. Balbinot, A. Fabbri, and A. Recati, 
 \href{https://doi.org/10.1140/epjd/e2009-00314-3}{Eur. Phys. J. D \textbf{56}, 391 (2010)}.

\bibitem{Jaskula} J. C. Jaskula, G. B. Partridge, M. Bonneau, R. Lopes, J. Ruaudel, D. Boiron, and C. I. Westbrook, 
 \href{https://doi.org/10.1103/PhysRevLett.109.220401}{Phys. Rev. Lett. \textbf{109}, 220401 (2012)}.

\bibitem{Koghee} S. Koghee and M. Wouters, 
 \href{https://doi.org/10.1103/PhysRevLett.112.036406}{Phys. Rev. Lett. \textbf{112}, 036406 (2014)}.

\bibitem{Busch} X. Busch, I. Carusotto, and R. Parentani, 
 \href{https://doi.org/10.1103/PhysRevA.89.043819}{Phys. Rev. A \textbf{89}, 043819(2014)}.

\bibitem{Saito} H. Saito and H. Hyuga, 
 \href{https://doi.org/10.1103/PhysRevA.78.033605}{Phys. Rev. A \textbf{78}, 033605 (2008)}.

\bibitem{Dodonov10} V. V. Dodonov and J. T. Mendonca, 
 \href{https://doi.org/10.1088/0031-8949/2014/T160/014008}{Phys. Scr. T \textbf{160}, 014008 (2014)}.






\bibitem{Aspelmeyer} M. Aspelmeyer, T. J. Kippenberg, and F. Marquardt, \href{https://doi.org/10.1103/RevModPhys.86.1391}{Rev. Mod. Phys. 86, 1391 (2014)}.

\bibitem{xsensing1} T. J. Kippenberg and K. J. Vahala, 
 \href{https://doi.org/10.1364/OE.15.017172}{Opt. Exp. \textbf{15}, 17172 (2007)}.

\bibitem{CQNCPRL} M. Tsang and C.M. Caves, 
 \href{https://doi.org/10.1103/PhysRevLett.105.123601}{Phys. Rev. Lett. \textbf{105} 123601 (2010)}.

\bibitem {CQNCPRX} M. Tsang and C.M. Caves, 
 \href{https://doi.org/10.1103/PhysRevX.2.031016}{Phys. Rev. X \textbf{2} 031016 (2012).}.

\bibitem{CQNCmeystre} F. Bariani, H. Seok, S. Singh, M. Vengalattore, and P. Meystre, 
 \href{https://doi.org/10.1103/PhysRevA.92.043817}{Phys. Rev. A \textbf{92}, 043817 (2015)}.

\bibitem{CQNCmaximilian} M. H. Wimmer, D. Steinmeyer, K. Hammerer, and M. Heurs, 
 \href{https://doi.org/10.1103/PhysRevA.89.053836}{Phys. Rev. A \textbf{89}, 053836 (2014)}.

\bibitem{aliNJP} Ali Motazedifard, F. Bemani, M. H. Naderi, R. Roknizadeh and D. Vitali, 
 \href{http://stacks.iop.org/1367-2630/18/i=7/a=073040}{New J. Phys. \textbf{18}, 073040 (2016) }.

\bibitem{complexCQNC} L. F. Buchmann, S. Schreppler, J. Kohler, N. Spethmann, and D. M. Stamper-Kurn, 
 \href{https://doi.org/10.1103/PhysRevLett.117.030801}{Phys. Rev. Lett. \textbf{117} 030801(2016) }.

\bibitem{ground state cooling} A. D. O’ Connell, M. Hofheinz, M. Ansmann, R. C. Bialczak, M. Lenander, E. Lucero, M. Neeley, D. Sank, H. Wang, M. Weides, J. Wenner, J. M. Martinis, and A. N. Cleland, 
 \href{https://doi.org/10.1038/nature08967}{Nature (London) \textbf{464}, 697 (2010)}.

\bibitem{Sideband cooling} J. D. Teufel, T. Donner, D. Li, J. W. Harlow, M. S. Allman, K. Cicak, A. J. Sirois, J. D. Whittaker, K. W. Lehnert, and R. W. Simmonds, 
 \href{https://doi.org/10.1038/nature10261}{Nature (London) \textbf{475}, 359 (2011)}.

\bibitem{Laser cooling} J. Chan, T. P. M. Alegre, A. H. Safavi-Naeini, J. T. Hill, A. Krause, S. Groblacher, M. Aspelmeyer, and O. Painter, 
 \href{https://doi.org/10.1038/nature10461}{Nature (London) \textbf{478}, 89 (2011)}.

\bibitem{Teufel} J. D. Teufel, T. Donner, D. Li, J.W. Harlow, M. S. Allman, K. Cicak, A. J. Sirois, J. D. Whittaker, K.W. Lehnert, and R.W. Simmonds, \href{https://doi.org/10.1038/nature10261}{Nature (London) \textbf{475}, 359 (2011)}.

\bibitem{Chan} J. Chan, T. P. M. Alegre, A. H. Safavi-Naeini, J. T. Hill, A. Krause, S. Grb\"{o}lacher, M. Aspelmeyer, and O. Painter, \href{https://doi.org/10.1038/nature10461}{Nature (London) \textbf{478}, 89 (2011)}.

\bibitem{Palomaki2} T. A. Palomaki, J. D. Teufel, R.W. Simmonds, and K.W. Lehnert,  \href{https://doi.org/10.1126/science.1244563}{Science \textbf{342}, 710 (2013)}.

\bibitem{Paternostro} M. Paternostro, D. Vitali, S. Gigan, M. S. Kim, C. Brukner, J. Eisert, and M. Aspelmeyer, \href{https://doi.org/10.1103/PhysRevLett.99.250401}{Phys. Rev. Lett. \textbf{99}, 250401 (2007)}

\bibitem{genes-entangelment} C Genes, A Mari, D Vitali, and P Tombesi,    \href{https://doi.org/10.1016/S1049-250X(09)57002-4}{Adv. At. Mol. Opt. Phys. \textbf{57}, 33 (2009)}.

\bibitem{Mari1} A. Mari, A. Farace, N. Didier, V. Giovannetti, and R. Fazio, 
 \href{https://doi.org/10.1103/PhysRevLett.111.103605}{Phys. Rev. Lett. \textbf{111}, 103605 (2013)}.

\bibitem{MianZhang} M. Zhang, G. S. Wiederhecker, S. Manipatruni, A. Barnard, P. McEuen, and M. Lipson, 
 \href{https://doi.org/10.1103/PhysRevLett.109.233906}{Phys. Rev. Lett. \textbf{109}, 233906 (2012)}.

\bibitem{Bagheri} M. Bagheri, M. Poot, L. Fan, F. Marquardt, and H. X. Tang, 
 \href{https://doi.org/10.1103/PhysRevLett.111.213902}{Phys. Rev. Lett. \textbf{111}, 213902 (2013)}.

\bibitem{Shlomi} K. Shlomi, D. Yuvaraj, I. Baskin, O. Suchoi, R. Winik, and E. Buks, 
 \href{https://doi.org/10.1103/PhysRevE.91.032910}{Phys. Rev. E \textbf{91}, 032910 (2015)}.

\bibitem{grebogi} Lei Ying, Ying-Cheng Lai, and Celso Grebogi, 
 \href{https://doi.org/10.1103/PhysRevA.90.053810}{Phys. Rev. A \textbf{90}, 053810 (2014)}.

\bibitem{foroud} F. Bemani, Ali Motazedifard, R. Roknizadeh, M. H. Naderi, and D. Vitali, 
 \href{https://journals.aps.org/pra/abstract/10.1103/PhysRevA.96.023805}{Phys. Rev. A \textbf{96}, 023805 (2017)}.

\bibitem{Borkje}A. Nunnenkamp, K. Borkje, and S. M. Girvin, 
 \href{https://doi.org/10.1103/PhysRevLett.107.063602}{Phys. Rev. Lett. \textbf{107}, 063602 (2011)}.

\bibitem{Hammerer} K. Jähne, C. Genes, K. Hammerer, M. Wallquist, E. S. Polzik, and P. Zoller, 
 \href{https://doi.org/10.1103/PhysRevA.79.063819}{Phys. Rev A. \textbf{79}, 063819 (2009)}.



\bibitem{dalafi1} A. Dalafi, M. H. Naderi, M. Soltanolkotabi, and Sh. Barzanjeh, 
 \href{https://doi.org/10.1103/PhysRevA.87.013417}{Phys. Rev. A \textbf{87} 013417 (2013) }.

\bibitem{dalafi2} A Dalafi, MH Naderi, M Soltanolkotabi, Sh Barzanjeh, 
 \href{https://doi.org/10.1088/0953-4075/46/23/235502}{J. Phys. B: At. Mol. Opt. Phys. \textbf{46}, 235502 (2013)}.

\bibitem{dalafi3} A. Dalafi and M. H. Naderi,
 \href{https://doi.org/10.1103/PhysRevA.94.063636}{Phys. Rev. A \textbf{94} 063636 (2016)}.

\bibitem{dalafi4} A. Dalafi and M. H. Naderi,
 \href{https://doi.org/10.1103/PhysRevA.95.043601}{Phys. Rev. A \textbf{95} 043601 (2017)}.








\bibitem{Salvatore} V. Macri, A. Ridolfo, O. Di Stefano, A. F. Kockum, F. Nori, and S. Salvatore, 
 \href{}{arXiv:\textbf{1706}.04134v1 [quant-ph](2017)}.

\bibitem{Mahajan} S. Mahajan, N. Aggarwal, T. Kumar, A. B. Bhattacherjee, and M. Mohan, 
 \href{https://doi.org/10.1139/cjp-2014-0255}{Can. J. Phys. \textbf{93},716 (2015)}.

\bibitem{Thompson} J. D. Thompson, B. M. Zwickl, A. M. Jayich, F. Marquardt, S. M. Girvin, and J. G. E. Harris, 
 \href{https://doi.org/10.1038/nature06715}{Nature (London) \textbf{452}, 72 (2008)}.


\bibitem{motazedi2} A. Motazedifard, M. H. Naderi, and R. Roknizadeh, 
 \href{https://doi.org/10.1364/JOSAB.99.099999}{J. Opt. Soc. Am. B \textbf{3}, 642(2017)}.








\bibitem{singlecavitymode} C. K. Law, 
 \href{https://doi.org/10.1103/PhysRevA.51.2537}{Phys. Rev. A \textbf{51}, 2537 (1995)}.

\bibitem{singlemechanicalmode} C. Genes, D. Vitali, and P. Tombesi, 
 \href{https://doi.org/10.1088/1367-2630/10/9/095009}{New J. Phys. \textbf{10}, 095009 (2008)}.

\bibitem{optomechanics with two phonon driving} B. A. Levitan, A. Metelmann, and A. A. Clerk, 
 \href{https://doi:10.1088/1367-2630/18/9/093014}{New. J. Phys. \textbf{18}, 093014 (2016)}.



\bibitem{Ritsch} C. Maschler and H. Ritsch, 
 \href{http://dx.doi.org/10.1016/j.optcom.2004.10.038}{Opt. Commun.  \textbf{243}, 145 (2004) }.

\bibitem{Domokos} Domokos P, Horak P and Ritsch H, 
 \href{http://dx.doi.org/10.1088/0953-4075/34/2/306}{J. Phys. B: At. Mol. Opt. Phys. \textbf{34} 187 (2001) }.

\bibitem{Morsch} O. Morsch and M. Oberthaler, 
 \href{https://doi.org/10.1103/RevModPhys.78.179}{Rev. Mod. Phys. \textbf{78}, 179(2006)}.

\bibitem{Cigar} J. M. Zhang, F. C. Cui, D. L. Zhou, and W. M. Liu,
 \href{https://doi.org/10.1103/PhysRevA.79.033401}{Phys. Rev. A \textbf{79} 033401 (2009) }.


\bibitem{Domokos2} G. Szirmai, D. Nagy, and P. Domokos, 
 \href{https://doi.org/10.1103/PhysRevA.81.043639}{ Phys. Rev. A \textbf{81} 043639(2010)}.

\bibitem{Nagy} D. Nagy, P. Domokos, A. Vukics, and H. Ritsch, 
 \href{https://doi.org/10.1140/epjd/e2009-00265-7}{Eur. Phys. J. D \textbf{55}, 659 (2009)}.



\bibitem{MeystreBEC}K. Zhang, W. Chen, M. Bhattacharya, and P. Meystre, 
 \href{https://doi.org/10.1103/PhysRevA.81.013802}{Phys. Rev. A \textbf{81} 013802 (2010)}.

\bibitem{dalafi qpt} A. Dalafi, M. H. Naderi, and M. Soltanolkotabi,
 \href{ https://doi.org/10.1088/0953-4075/48/11/115507}{J. Phys. B \textbf{48}, 115507 (2015)}.
 
 
\bibitem{MeystreBEC2} S. K. Steinke and P. Meystre, 
 \href{https://doi.org/10.1103/PhysRevA.84.023834}{ Phys. Rev. A \textbf{84} 023834 (2011)}.

\bibitem{MeystreBEC3} W. Chen, D. S. Goldbaum, M. Bhattacharya, and P. Meystre, 
 \href{https://doi.org/10.1103/PhysRevA.81.053833}{Phys. Rev. A \textbf{81} 053833 (2010)}.


\bibitem{Gardiner}C. W. Gardiner and P. Zoller, \textit{Quantum Noise} (Springer, Berlin, 2000).


\bibitem{vitali noise membrane} V. Giovannetti and D. Vitali, 
 \href{https://doi.org/10.1103/PhysRevA.63.023812}{Phys. Rev. A \textbf{63}, 023812 (2001)}.


\bibitem{Routh} E. X. DeJesus and C. Kaufman, 
\href{https://doi.org/10.1103/PhysRevA.35.5288}{Phys. Rev. A. \textbf{35}, 5288 (1987)}.

\bibitem{JLiNJP} J. Li, I. M. Haghighi, N. Malossi, S. Zippilli1, and D. Vitali, 
 \href{https://doi:10.1088/1367-2630/17/10/103037}{New. J. Phys. \textbf{17}, 103037 (2015)}.

\bibitem{JLiPRA} J. Li, G. Li, S. Zippilli, D. Vitali, and T. Zhang, 
 \href{https://doi.org/10.1103/PhysRevA.95.043819}{Phys. Rev. A \textbf{95}, 043819 (2017)}.

\bibitem{dalafi5} A. Dalafi, M. H. Naderi and M. Soltanolketabi,
 \href{http://dx.doi.org/10.1080/09500340.2014.935818}{J. Mod. Opt. \textbf{61} 1387 (2014)}.

\bibitem{antiDCE-work} A. V. Dodonov, D. Valente, and T. Werlang, 
 \href{https://doi.org/10.1103/PhysRevA.96.012501}{Phys. Rev. A \textbf{96}, 012501 (2017)}.

\bibitem{antiDCE} L. C. Monteiroa and A. V. Dodonov,
 \href{https://doi.org/10.1016/j.physleta.2016.02.031}{Phys. Lett. A. \textbf{380}, 1542 (2016)}.


\end{thebibliography}
\end{document}